\documentclass{aa}
\usepackage{natbib}
\bibpunct{(}{)}{;}{a}{}{,} 
\usepackage[utf8]{inputenc}

\usepackage{multicol}
\usepackage{booktabs}
\usepackage{caption}

\usepackage{graphicx}
\usepackage[varg]{txfonts}

\usepackage{multicol}
\usepackage{lscape}

\title{Radio observations of massive stars in the Galactic centre: The Quintuplet cluster}
\author{Gallego-Calvente\inst{1}\fnmsep\thanks{\email{gallego@iaa.es}}, A.~T.
        \and
         Sch\"odel\inst{1}, R.
        \and
         Alberdi \inst{1}, A.
        \and
         Najarro\inst{2}, F.
        \and
         Yusef-Zadeh \inst{3}, F.
        \and
         Shahzamanian\inst{1}, B.
        \and
         Nogueras-Lara\inst{4}, F.
        }
         
\institute{Instituto de Astrof\'isica de Andaluc\'ia (IAA-CSIC), Glorieta de la Astronom\'ia s/n,
              18008 Granada, Spain e-mail: gallego@iaa.es
         \and 
             Centro de Astrobiolog\'ia (CSIC/INTA), Ctra. de Ajalvir Km. 4, 28850 Torrej\'on de Ardoz, Madrid, Spain
         \and 
	         CIERA, Department of Physics and Astronomy Northwestern University, Evanston, IL 60208, USA
         \and 
         	 Max-Planck-Institut f\"ur Astronomie, K\"onigstuhl 17, Heidelberg, D-69 117, Germany
             }
\date{}

\abstract{
   We present high-angular-resolution radio continuum observations of the Quintuplet cluster, one of the most emblematic massive clusters in the Galactic centre. Data were acquired in two epochs and at 6 and 10\,GHz with the Karl J. Jansky Very Large Array. With this work, we have quadrupled the number of known radio stars in the cluster. Nineteen of them have spectral indices consistent with thermal emission from ionised stellar winds, five are consistent with colliding wind binaries, two are ambiguous cases, and one was only detected in a single band. Regarding variability, remarkably we find a significantly higher fraction of variable stars in the Quintuplet cluster ($\sim$\,$30\%$) than in the Arches cluster (< $15\%$), probably due to the older age of the Quintuplet cluster. Our determined stellar wind mass-loss rates are in good agreement with theoretical models. Finally, we show that the radio luminosity function can be used as a tool to constrain the age and the mass function of a cluster.
}\keywords{}

\begin{document}

\maketitle

\section{Introduction}

The Quintuplet cluster, along with the Arches and the Central Parsec clusters, is a young, massive stellar cluster in the Galactic centre (GC). Separated just a few arcminutes from the Arches cluster, the Quintuplet cluster is located at a projected distance of approximately 30\,pc (or 12.5\,\arcmin\, in angular distance) to the northeast of Sagittarius A* (Sgr\,A*), the black hole at the centre of our Galaxy. Given its position at the GC, Quintuplet can only be observed (along with at radio frequencies) at infrared wavelengths, which poses a difficulty in identifying its stellar content, because intrinsic stellar colours are small in the infrared and reddening and the related uncertainty will dominate the observed colours \citep{Nogueras-Lara:2018aa, Nogueras-Lara:2021aa}.

\begin{figure*}[t!]
    \centering
    \includegraphics[width=.8\textwidth]{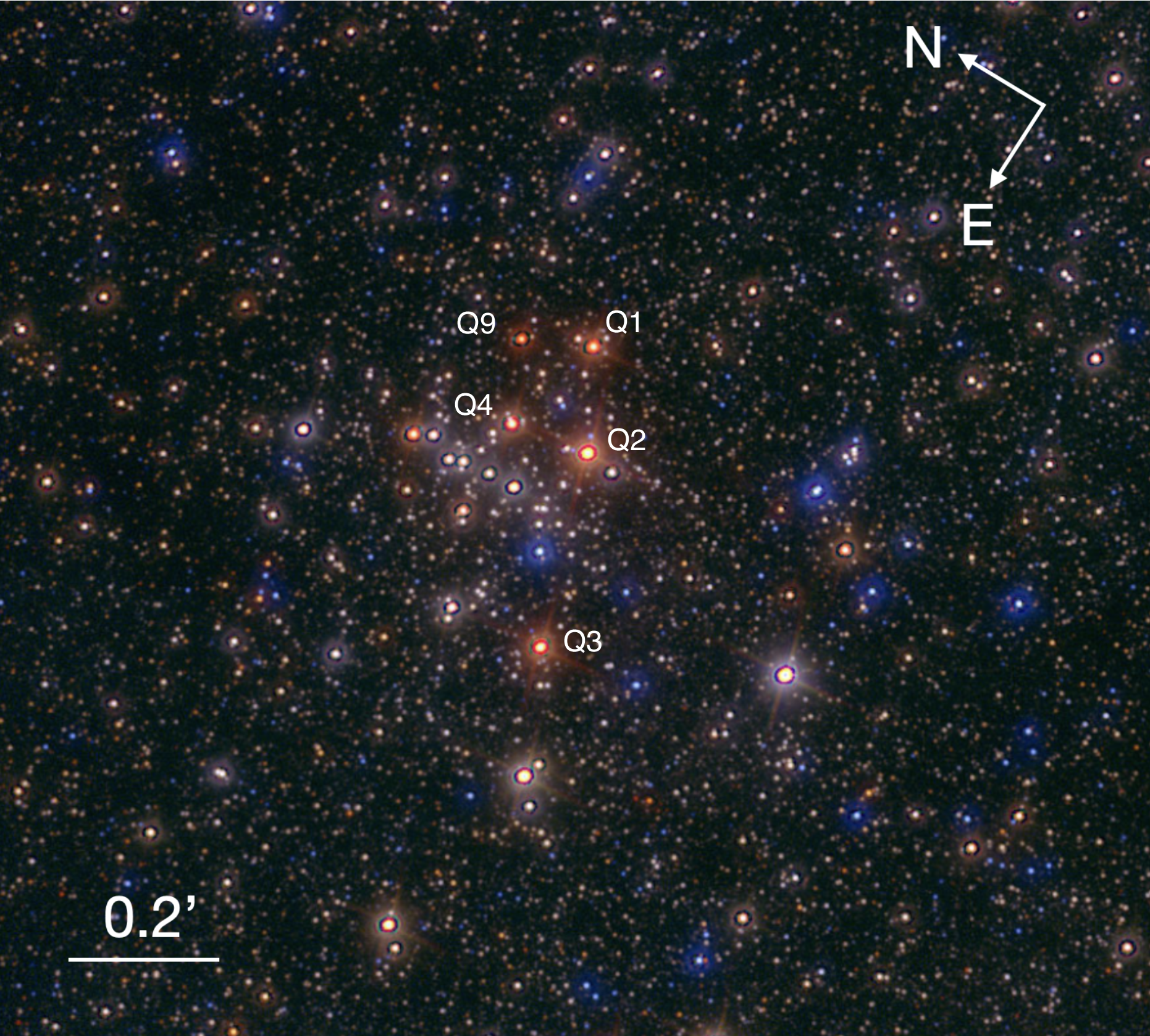}
    \caption{JHKs false colour image of the Quintuplet cluster from the GALACTICNUCLEUS survey. Credits: \citet{Nogueras-Lara:2018aa,Nogueras-Lara:2019ab}.}
    \label{fig:quintuplet}
\end{figure*}

The Quintuplet cluster owes its name to five prominent infrared-bright sources, later named Q1, Q2, Q3, Q4 and Q9 by \citet{Figer:1999ab} (see Fig.\,\ref{fig:quintuplet}). It has a similar mass as the Arches cluster, about $1-2\times10^{4}$\,M$_{\odot}$ \citep{Figer:1999ab,Rui:2019ch}, but is considered to be roughly 1\,Myr older than the former \citep[Arches: 2.5$-$4\,Myr, Quintuplet: 3$-$5\,Myr;][]{Figer:1999ab,Clark:2018yp,Clark:2018aa,Schneider:2014vn}, which is reflected in its larger core radius -- probably a sign of its ongoing dissolution in the GC tidal field \citep{Rui:2019ch} -- and its content of more evolved stars than the Arches cluster \citep{Clark:2018aa}. The extended nature of the Quintuplet not only poses a problem with respect to confusion with the (mostly old) stars of the nuclear stellar disk \citep{Nogueras-Lara:2019gm, Launhardt:2002aa}, in which it is embedded, but there may also be problems to interlopers of young stars that have formed in the vicinity of the Quintuplet, but at a somewhat different time \citep{Clark:2018aa}.


Massive young stellar clusters are of great interest to study the evolution of the most massive stars, with masses in excess of a few tens of solar masses. The observable number of such stars in the Milky Way is small because they are very rare due to the stellar initial mass function that is a steep power-law of mass \citep{Hosek:2019vn} and due to the very short lifetimes (at most a few Myr) of these stars. The latter fact means that massive stars are typically not only located at distances of many kiloparsecs from Earth, but also frequently still enshrouded in dusty molecular clouds and therefore highly extinguished. As two of the most massive young clusters in the Milky Way, the Arches and the Quintuplet have been the object of intense study over the past three decades \citep[][among others]{Figer:1999ab, Najarro:2009aa, Najarro:2017aa, Clark:2018aa, Liermann:2009aa, Liermann:2010aa, Liermann:2012aa, Hosek:2019vn}, driven by ever improving infrared imaging and spectroscopy instruments. Imaging, spectroscopic and proper motion measurements in the infrared have provided us with our current ideas of mass, age, and metallicity of these clusters as the spectral types and mass of their $\sim$100 most massive stars, as described in the references cited above.

Highly complementary information on the properties of massive young stars can be obtained from radio  observations. Radio continuum emission arises in the ionised winds of massive stars \citep[see, e.g.,][]{Leitherer:1997yf,Gudel:2002rt,Umana:2015yq} and can provide essential information on mass-loss rates through stellar winds. \citet{Lang:2005aa}  carried out a multi-frequency, multi-configuration, and multi-epoch study of the Arches and Quintuplet clusters with the Very Large Array (VLA). They detected ten more or less compact radio sources. The majority of them had rising spectral indices as expected for young massive stars with powerful stellar winds and a few of them had clear near-infrared counterparts. 

In this work, we study the Quintuplet with the Karl G. Janksy Very Large Array (JVLA) taking advantage of its significantly increased sensitivity in comparison with the old VLA, as we did for the Arches cluster aiming to pick up the thermal and non-thermal emission from the ionised gas in the outer wind regions of young, massive stars \citep[see ][]{Gallego-Calvente:2021yl}.
 
\section{Observations and imaging \label{sec4:obs}}

We observed the radio continuum emission from the Quintuplet cluster using the National Radio Astronomy Observatory (NRAO)\footnote{The NRAO is a facility of the National Science Foundation (NSF) operated under cooperative agreement by Associated Universities, Inc.} JVLA in 2016 and 2018. We acquired data in two bands (C and X), with X-band observations in three epochs, as detailed in Table\,\ref{table:quintuplet_obs}.
The phase centre was taken at the position $\alpha$, $\delta_{\textrm{(J2000)}}$ = 17$^{\textrm{h}}$\,46$^{\textrm{m}}$ \,15.26$^{\textrm{s}}$, $-$28\degr\,49\arcmin\,33.0\arcsec\,. All  observations were carried out in the A configuration to achieve the highest angular resolution. This configuration also helped us to filter out part of the extended emission that surrounds the Quintuplet cluster whose center is located at a few arcminutes to the southeast of the Sickle H\,II region (G0.18$-$0.04) and at approximately 10\arcsec due north of the Pistol (G0.15$-$0.05) H\,II region. 

\begin{table}[t]
  \centering
	\caption[Quintuplet cluster: Observational properties]{Summary of observations}
    \footnotesize
    \setlength{\tabcolsep}{4pt}
	\begin{tabular}{c c c c}
	    \hline \hline
		Observation & Band \footnotesize{$^{\mathrm{a}}$} & JVLA            & On source time \\
		date        &                                     & configuration   & (minutes) \\
		\noalign{\smallskip}
		\hline
		\noalign{\smallskip}
	        Oct 04, 2016  & X & A & 55 \\
            Oct 27, 2016  & X & A & 55 \\
            Mar 24, 2018  & X & A & 55 \\
            Jun 10, 2018  & C & A & 74 \\
        \hline
	\end{tabular}
    \begin{list}{}{}\itemsep1pt \parskip0pt \parsep0pt \footnotesize
        \item[$^{\mathrm{a}}$] Frequency range of $8-12$\,GHz for X-band and $4-8$\,GHz for C-band. Therefore, the total bandwidth was 4\,GHz on each band. The number of spectral windows was 32 and the number of channels 64 in both cases.
    \end{list}
	\label{table:quintuplet_obs}
\end{table}

J1744$-$3116 was used as phase calibrator and J1331+305 (3C286) as band-pass and flux density calibrator at all frequencies. 
The raw data were processed automatically performing an initial flagging and calibration through the JVLA calibration pipeline. Extra flagging was necessary to remove lost or corrupted data. The entire cluster was visible within the Field of View (FoV) of the observations.

A very bright source was present in the FoV in both bands. This source is located at $\alpha$, $\delta_{\textrm{(J2000)}}$ = 17$^{\textrm{h}}$\,46$^{\textrm{m}}$ \,21.04$^{\textrm{s}}$, $-28$\degr\,50\arcmin\,03.11\arcsec.\: We identified it as SgrA-N3 using the VizieR Catalogue Service\footnote{http://vizier.u-strasbg.fr/}. We carried out data reduction with the Common Astronomy Software Applications package (CASA) developed by an international consortium of scientists\footnote{Scientists based at the National Radio Astronomical Observatory (NRAO), the European Southern Observatory (ESO), the National Astronomical Observatory of Japan (NAOJ), the Academia Sinica Institute of Astronomy and Astrophysics (ASIAA), the CSIRO division for Astronomy and Space Science (CASS), and the Netherlands Institute for Radio Astronomy (ASTRON) under the guidance of NRAO.} as follows.

We used the task {\tt tclean} to create a model for the intense source in order to, later on, self-calibrate on this source. The $u-v$ range was constrained to frequencies $>200$\,k$\lambda$ to filter out some of the extended emission near and around the target field while detecting the compact sources. These two steps, that is model generation and calibration of the time-dependent antenna-based gains on the bright source, were repeated several times, thereby iteratively changing the parameter that controls the solution interval, {\tt solint}, in the {\tt gaincal} task. An improvement of 15\% in the dynamic range was achieved through this iterative approach. We inserted the final model by means of the task {\tt ft} from CASA. Since the diffraction pattern residuals of the bright ($\sim$40\,mJy) SgrA-N3 source interfered with our aim to detect faint sources, we subtracted it in the Fourier plane. Subsequently, we cleaned our new visibility data set, stored in a new table as a {\tt Measurement Set} ({\tt MS}), using {\tt tclean} in interactive mode. We chose the interactive mode because the subtraction of SgrA-N3 was not completely perfect and the restriction of the $u-v$ range did not entirely remove the extended emission.

 We atempted to improve the images in all epochs through testing the more advanced form of imaging\; {\tt multi-scale\,clean}, that distinguishes emitting features with angular scales between point sources and extended emission. Its use did not improve the quality of the final images. Additionally, we probed different weighting schemes to correct for visibility sampling effects. The natural weighting scheme resulted in the images with the highest signal-to-noise (S/N). The gain parameter in the clean algorithm was set to 0.05. With this procedure we reached an off-source rms noise level of 4.3\,$\mu$Jy\,beam$^{-1}$ in the 2016 X-band image in the central regions of the FoV. The off-source thermal noise in the 2018 epochs was $5.7$ and 6.6\,$\mu$Jy\,beam$^{-1}$ for X- and C-bands, respectively.

All the images were primary beam corrected to account for the change in sensitivity across the primary beam. Table\,\ref{table:quintuplet_prop} summarises the properties of the final images.

\begin{table*}[th!]
	\caption{Properties of the images}
	\begin{tabular}{c c c c r c c}
	\hline \hline
	\centering
          Epoch & Band & Frequency \footnotesize{$^{\mathrm{a}}$} & Synthesised beam & \multicolumn{1}{r}{P. A. \footnotesize{$^{\mathrm{b}}$}}  & rms noise \footnotesize{$^{\mathrm{c}}$}  & $(u,v)$ cut-off   \\
                &      & (GHz)                       & (arcsec $\times$ arcsec)             & \multicolumn{1}{c}{(degrees)}               & ($\mu$Jy\,beam$^{-1}$)              & (k$\lambda$)  \\
          \noalign{\smallskip}
          \hline
          \noalign{\smallskip}
          2016 & X & 10.0  &  $0.47 \times 0.15$  &    25.29    &  4.3  & 200   \\
          2018 & X & 10.0  &  $0.41 \times 0.16$  & $-$6.84     &  5.7  & 200   \\
          2018 & C &  6.0  &  $0.62 \times 0.21$  &    19.97    &  6.6  & 200   \\
    \hline
	\end{tabular}
    \begin{list}{}{}\itemsep1pt \parskip0pt \parsep0pt \footnotesize
        \item[$^{\mathrm{a}}$] Representative frequency, in Gigahertz.
        \item[$^{\mathrm{b}}$] The position angle (P. A.) of the fitted major axis for the synthesised beam, in degrees.
        \item[$^{\mathrm{c}}$] Off-source root mean square noise level reached.
    \end{list}
	 \label{table:quintuplet_prop}
\end{table*}


\section{Results \label{sec4:results}}

\subsection{Point source detection and flux density \label{sec4:alphas}}

Point sources were selected interactively in the cleaning procedure, as we specified in section\,\ref{sec4:obs}, restricting ourselves to those at above five times the off-source rms noise level.\\
We used the image-plane component fitting ({\tt imfit}) task from CASA that takes into account the quality of the fit and each image rms, to estimate the positions of the maxima, the total flux densities, and the errors of these values over the final primary-beam-corrected images for all bands in all epochs.
We have determined the uncertainties in the positions of the detected sources by adding in quadrature to the formal error of the fit, 0.5\,$\theta$/SNR \citep{reid:1988aa} a systematic error of 0.05\arcsec \citep{dzib:2017aa}. In the formal error, $\theta$ is the source size convolved with the beam and SNR the S/N.
The systematic error addresses the thermal noise and uncertainties introduced by the phase calibration process. \\
Likewise, we considered the percentage in the calibration error of the peak flux densities at the observed frequencies \citep{perley:2013aa} and the factor that considers whether a source is resolved or unresolved to evaluate the flux-density uncertainties.
    
\begin{figure}[htp!]
    \centering
    \includegraphics[width=\columnwidth]{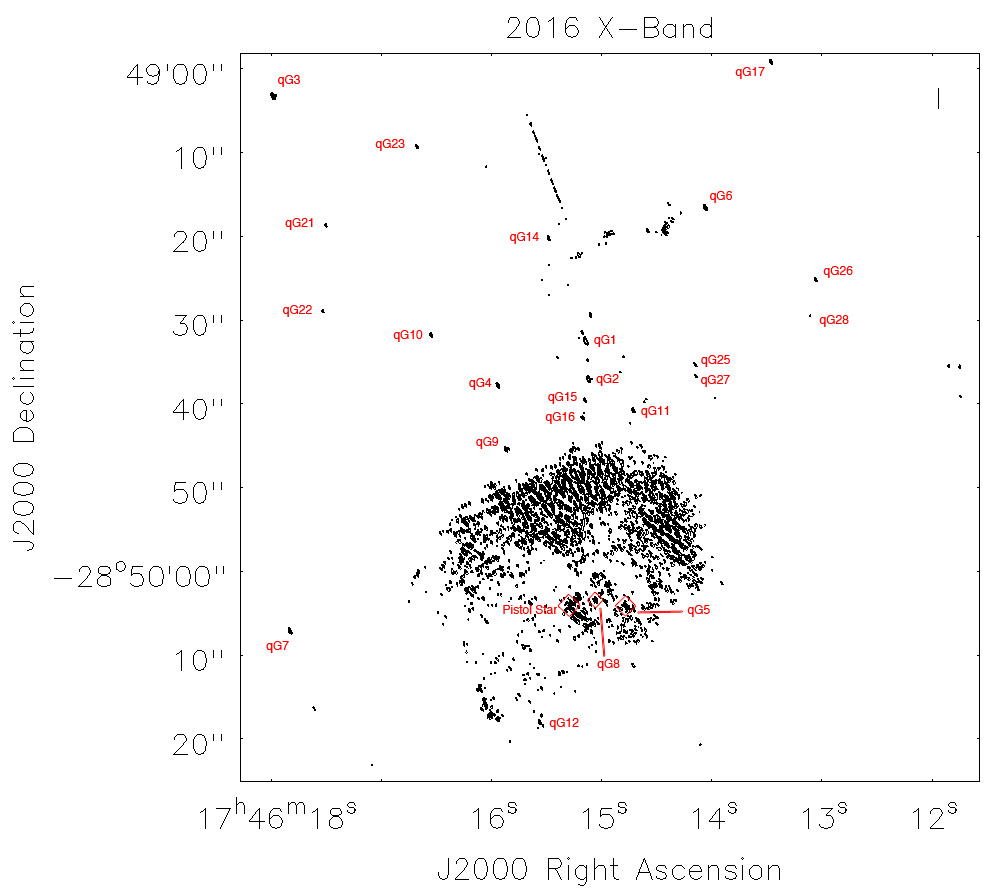}
    \includegraphics[width=\columnwidth]{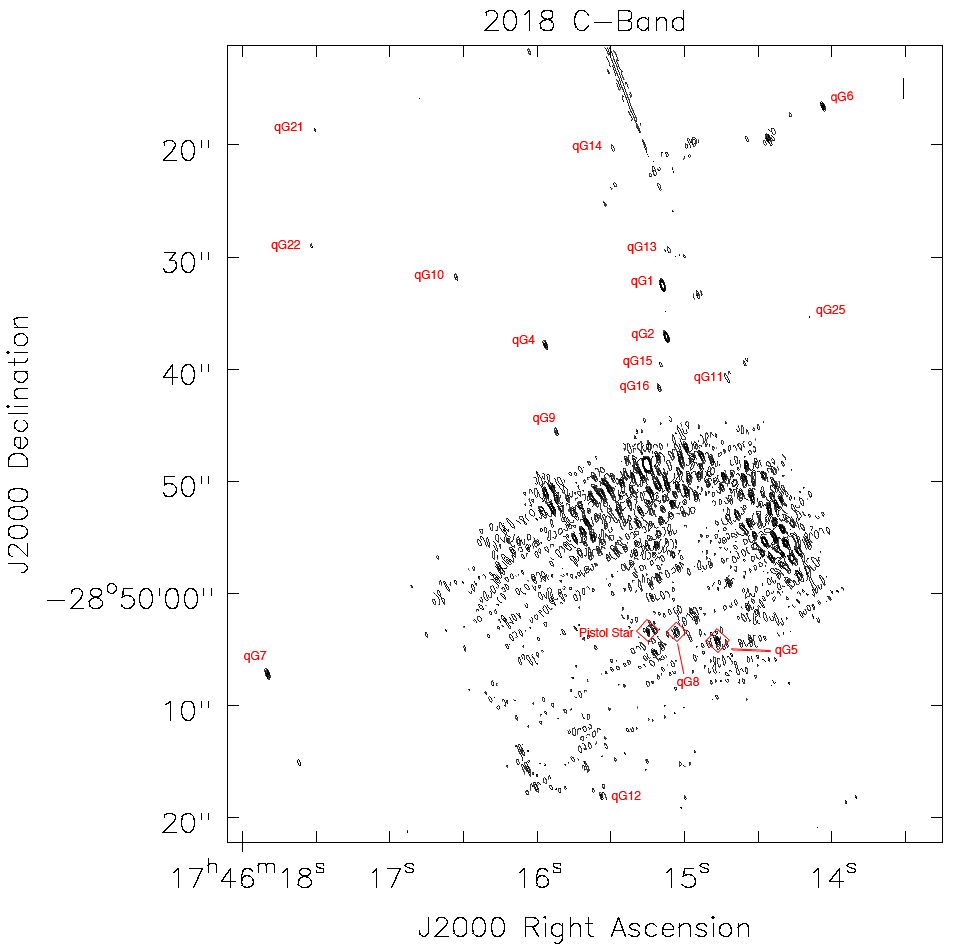}
    \caption[Quintuplet cluster: 2016 X- and 2018 C- bands closeup images]{Top: Closeup onto the Quintuplet cluster from the 2016 X-band image corrected for primary beam attenuation. The clean beam is 0.47$\arcsec \times$\,0.15$\arcsec$, P.A. = 25.29$\degr$. The off-source rms noise level is 4.3\,$\mu$Jy\,beam$^{-1}$. The contour levels represent -1, 1, 2, 3, 4, 5, and 6 times 5\,$\sigma$.
    Bottom: Closeup onto the Quintuplet cluster showing most of the detected sources from the 2018 C-band image corrected for primary beam attenuation. The resolution is 0.41$\arcsec \times$ 0.16$\arcsec$, P.A. = $-$6.84$\degr$. The off-source rms noise level is 5.7\,$\mu$Jy\,beam$^{-1}$. The contour levels represent -1, 1, 2, 3, 4, 5, and 6 times 5\,$\sigma$.}
    \label{fig:X-Quintuplet}
\end{figure}

This process led us to detect 29 and 28 point sources above 5\,$\sigma$ at 10\,GHz and 6\,GHz, respectively. 
Figure\,\ref{fig:X-Quintuplet} shows a closeup onto the 2016 X-band image and a closeup onto the 2018 C-band image of the cluster, both images corrected for primary beam attenuation, with radio stars labelled. We note that some of the detected point sources lie outside of the FoV shown in the figure. 

\begin{figure*}[th!]
    \centering
    \includegraphics[width=0.8\textwidth]{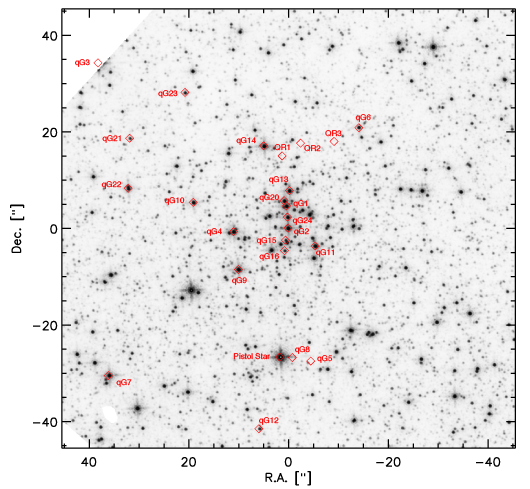}
    \caption{HST/WFC3 F153M image of the Quintuplet cluster with identified radio stars labelled.}
    \label{fig:Quintuplet_NIR}
\end{figure*}

The 2016 X-band image turned out to be the image with the highest S/N, so it provides the most complete list of point sources. Besides the rms noise, there is systematic noise present in the radio image, caused by remnant side lobes of SgrA-N3 that could not be fully removed from the image and some diffuse flux still present in the $u-v$ range considered for the image reconstruction. Also, the Quintuplet region is full of ionised gas in clouds of variable compactness. The systematic noise and the ionised gas can give rise to numerous spurious detections, even at 5\,$\sigma$ above the rms noise.  Therefore, we compared the positions of detected radio point  sources in our 2016 X-band image with the positions of stars on an HST/WFC3 F153M image of the Quintuplet cluster \citep[][]{Rui:2019ch}, downloaded from the HST archive\footnote{Based on observations made with the NASA/ESA Hubble Space Telescope, obtained from the data archive at the Space Telescope Science Institute. STScI is operated by the Association of Universities for Research in Astronomy, Inc. under NASA contract NAS 5-26555.}. All radio stars must necessarily be very bright infrared sources (massive, young stars) that are detected at S/N ratios exceeding a few times 100. Only sources coincident within 0.05$\arcsec$ with the position of stars detected in the NIR were accepted as real. The tagged HST/WFC3 F153M image is shown in Fig.\,\ref{fig:Quintuplet_NIR}.

Finally, we matched our sources  with the ones provided by \citep[][Tables A.1 and A.2]{Clark:2018aa}, based on HST/NICMOS+WFC3 photometry and VLT/SINFONI+ KMOS spectroscopy for $\sim$\,100 and 71 cluster members, respectively. In this way we could obtain the spectral type of the radio stars.

\begin{figure*}[ht!]
    \centering
    \includegraphics[width=0.7\textwidth]{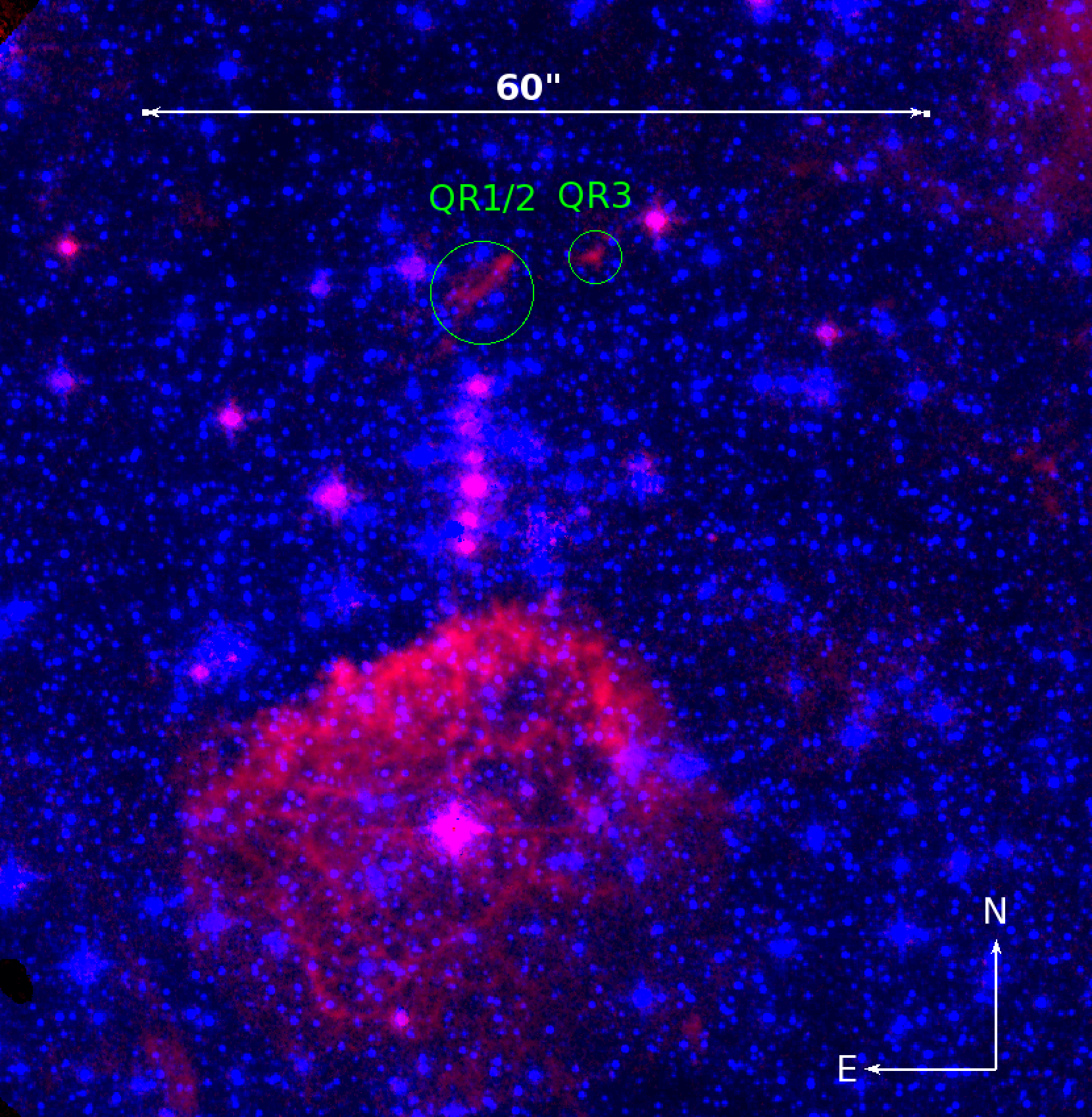}
    \caption[Composition of HST/WFC3 F153M and Paschen $\alpha$ images]{Composition of the HST/WFC3 F153M image from HST archive with a Paschen $\alpha$ image by \citet{Dong:2011aa}. BLue: HST WFC3 F153M; Red: HST NIC3 Paschen $\alpha$. QR1$-$3 sources by \citet{Lang:2005aa} are encircled.}
    \label{fig:nir_paalpha}
\end{figure*}

\citealt{Lang:2005aa} reported nine compact radio sources and the Pistol Star. They referred to the former as QR1$-$QR9. QR1$-$3 were interpreted as the first detections of embedded massive stars in the Quintuplet cluster. As we can observe in Fig.\,\ref{fig:Quintuplet_NIR}, those sources have no bright, stellar NIR counterparts, as was clearly indicated by \citet{Lang:2005aa} as well. To further investigate these sources, we superposed the HST/WFC3 F153M image from HST archive with a Paschen $\alpha$ image by \citet{Dong:2011aa}, see Figure\,\ref{fig:nir_paalpha}. From this figure we can see that QR1$-$3 are not stellar sources but related to ionised gas clouds that have similar appearance as many other such objects in this region.

Considering this finding, we have renamed the Quintuplet cluster members to reassign a number to the detected radio sources from \#1 to \#29. We do not use the prefix QR used by \citet{Lang:2005aa} to avoid confusion between the old QR1$-$3 and the new 1$-$3 numbered members. We confirm seven sources reported by \citet{Lang:2005aa} as radio stars (QR4, QR5, QR6, QR7, QR8, QR9, Pistol Star). With a total of 29 sources reported here, this work quadruples the number of known radio stars associated with the Quintuplet cluster, with the faintest radio source having a flux density of 30\,$\pm$\,7\,$\mu$Jy at X-Band.

Table\,\ref{table:quintuplet_results} shows the new and the old nomenclature, the identified NIR counterparts as listed in \citet{Clark:2018aa}, determined positions, the X- and C-band flux densities for all radio sources measured in our 2016 and 2018 data sets or the upper limits of the undetected sources, and the spectral classification according to \citet{Clark:2018aa}.

\subsection{Spectral indices}
 We calculated the spectral indices (or their limits) of the radio stars, as well as the corresponding uncertainties using their measured X- and C-band fluxes. We proceeded as described in  \citet{Gallego-Calvente:2021yl}. The spectral indices are listed in column\,8 of Table\,\ref{table:quintuplet_results}.

\begin{sidewaystable*}[tph!]
    \caption[Quintuplet cluster: JVLA flux densities of the compact sources]{JVLA flux densities of the compact sources of the Quintuplet cluster from these observations.}
    \footnotesize
    \setlength{\tabcolsep}{3.1pt} 
    \begin{tabular}{l l l c c c c r l}
	    \hline \hline
		\multicolumn{1}{ c }{Source} & \multicolumn{1}{ c }{Near-IR} & \multicolumn{2} { c } {Positions in X band (J2000.0) \footnotesize{$^{\mathrm{c}}$}} & \multicolumn{3} { c } {Flux density (mJy) \footnotesize{$^{\mathrm{d}}$}} & \multicolumn{1}{ c }{$\alpha$} & \multicolumn{1}{ c }{Spectral} \\ \cmidrule(lr{.5em}){3-4}  \cmidrule(lr{.5em}){5-7}
		\multicolumn{1}{ c }{name}   & \multicolumn{1}{ c }{counterpart \footnotesize{$^{\mathrm{b}}$}}  & \multicolumn{1}{ c }{R.A.} & Dec. & X-Band (2016) & X-Band (2018) & C-band (2018) & & \multicolumn{1}{ c }{type \footnotesize{$^{\mathrm{f}}$}} \\
		\hline
        qG1 (QR6 \footnotesize{$^{\mathrm{a}}$})    &  qF257, LHO096    & 17 46 15.14 $\pm$ 0.01  & $-$28 49 32.46 $\pm$ 0.01  & $ 1.1   \pm 0.1$   & $0.49  \pm 0.05$  &  $0.45  \pm 0.05$  &    $0.2 \pm 0.3$  & B1-2 Ia$^{+}$ \\
        qG2 (QR5 \footnotesize{$^{\mathrm{a}}$})    &  qF241, LHO071    & 17 46 15.12 $\pm$ 0.01  & $-$28 49 37.00 $\pm$ 0.01  & $ 0.50  \pm 0.05$  & $0.52  \pm 0.05$  &  $0.34  \pm 0.03$  &    $0.8 \pm 0.3$  & WN11h  \\
        qG3 (QR9 \footnotesize{$^{\mathrm{a}}$})    &  qF362            & 17 46 17.98 $\pm$ 0.01  & $-$28 49 03.20 $\pm$ 0.01  & $ 0.40  \pm 0.04$  & $0.50  \pm 0.05$  &  $0.17  \pm 0.02$  &    $2.1 \pm 0.3$  & LBV   \\
        Pistol Star \footnotesize{$^{\mathrm{a}}$}  &  qF134            & 17 46 15.24 $\pm$ 0.01  & $-$28 50 03.37 $\pm$ 0.01  & $ 0.26  \pm 0.03$  & $0.34  \pm 0.03$  &  $0.10  \pm 0.01$  &    $2.4 \pm 0.3$  & LBV   \\
        qG4                                         &  qF240, LHO067    & 17 46 15.94 $\pm$ 0.01  & $-$28 49 37.73 $\pm$ 0.01  & $ 0.23  \pm 0.02$  & $0.21  \pm 0.02$  &  $0.14  \pm 0.02$  &    $0.8 \pm 0.3$  & WN10h \\
        qG5                                         &                   & 17 46 14.77 $\pm$ 0.01  & $-$28 50 04.20 $\pm$ 0.01  & $ 0.23  \pm 0.02$  & $0.19  \pm 0.02$  &  $0.30  \pm 0.03$  &   $-0.9 \pm 0.3$  & <B0 I  \\
        qG6 (QR8 \footnotesize{$^{\mathrm{a}}$})    &  qF320, LHO158    & 17 46 14.05 $\pm$ 0.01  & $-$28 49 16.47 $\pm$ 0.01  & $ 0.22  \pm 0.02$  & $0.22  \pm 0.02$  &  $0.16  \pm 0.02$  &    $0.6 \pm 0.3$  & WN9h  \\
        qG7                                         &                   & 17 46 17.83 $\pm$ 0.01  & $-$28 50 07.16 $\pm$ 0.01  & $ 0.17  \pm 0.02$  & $0.18  \pm 0.02$  &  $0.20  \pm 0.02$  &   $-0.2 \pm 0.3$  & O7-8 Ia$^{+}$/WNLh  \\
        qG8                                         &                   & 17 46 15.05 $\pm$ 0.01  & $-$28 50 03.47 $\pm$ 0.01  & $ 0.15  \pm 0.02$  & $0.23  \pm 0.02$  &  $0.20  \pm 0.02$  &    $0.3 \pm 0.3$  & WC8-9(d? + OB?)     \\
        qG9                                         &  qF211, LHO019    & 17 46 15.87 $\pm$ 0.01  & $-$28 49 45.46 $\pm$ 0.01  & $ 0.12  \pm 0.01$  & $0.10  \pm 0.01$  &  $0.09  \pm 0.01$  &    $0.2 \pm 0.3$  & WC9d (+OB) \\
        qG10                                        &  qF256, LHO099    & 17 46 16.55 $\pm$ 0.01  & $-$28 49 31.73 $\pm$ 0.01  & $ 0.12  \pm 0.01$  & $0.09  \pm 0.01$  &  $0.060 \pm 0.009$ &    $0.8 \pm 0.4$  & WN8-9ha \\
        qG11 (QR7 \footnotesize{$^{\mathrm{a}}$})   &  qF231, LHO042    & 17 46 14.71 $\pm$ 0.01  & $-$28 49 40.65 $\pm$ 0.01  & $ 0.10  \pm 0.01$  & $0.070 \pm 0.009$ &  $0.09  \pm 0.01$  &   $-0.5 \pm 0.3$  & WC9d(+OB) \\
        qG12                                        &                   & 17 46 15.56 $\pm$ 0.01  & $-$28 50 18.05 $\pm$ 0.01  & $ 0.10  \pm 0.01$  & $0.17  \pm 0.02$  &  $0.07  \pm 0.01$  &    $1.7 \pm 0.4$  & O7-8 Ia \\
        qG13 (QR4 \footnotesize{$^{\mathrm{a}}$})   &  qF270S, LHO110   & 17 46 15.10 $\pm$ 0.01  & $-$28 49 29.31 $\pm$ 0.01  & $ 0.09  \pm 0.01$  & $0.08  \pm 0.01$  &  $0.060 \pm 0.009$ &    $0.6 \pm 0.4$  & B1-2 Ia$^{+}$/WNLh \\
        qG14                                        &  qF307A, LHO146   & 17 46 15.48 $\pm$ 0.01  & $-$28 49 20.17 $\pm$ 0.01  & $ 0.09  \pm 0.01$  & $0.09  \pm 0.01$  &  $0.07  \pm 0.01$  &    $0.5 \pm 0.4$  & B1-2 Ia$^{+}$ \\
        qG15                                        &  qF235N, LHO047   & 17 46 15.15 $\pm$ 0.02  & $-$28 49 39.47 $\pm$ 0.02  & $ 0.080 \pm 0.009$ & $0.08  \pm 0.01$  &  $0.050 \pm 0.009$ &    $0.9 \pm 0.4$  & WC8(d? +OB?) \\
        qG16                                        &  qF235S, LHO034   & 17 46 15.17 $\pm$ 0.01  & $-$28 49 41.61 $\pm$ 0.01  & $ 0.080 \pm 0.009$ & $0.10  \pm 0.01$  &  $0.09  \pm 0.01$  &    $0.2 \pm 0.3$  & WC8(d? +OB?) \\
        qG17                                        &  qF381            & 17 46 13.45 $\pm$ 0.01  & $-$28 48 59.12 $\pm$ 0.01  & $ 0.074 \pm 0.008$ & $0.07  \pm 0.03$  &  $0.040 \pm 0.008$ &    $1.1 \pm 0.9$  & B0-1 Ia$^{+}$/WNLh \\
        qG18                                        &  qF353E           & 17 46 11.13 $\pm$ 0.01  & $-$28 49 05.85 $\pm$ 0.01  & $ 0.072 \pm 0.008$ & $0.07  \pm 0.03$  &  $0.030 \pm 0.008$ &    $2   \pm 1$    & WN6 \\
        qG19                                        &                   & 17 46 16.50 $\pm$ 0.01  & $-$28 48 44.08 $\pm$ 0.01  & $ 0.060 \pm 0.007$ & $0.060 \pm 0.009$ &  $0.08  \pm 0.01$  &   $-0.6 \pm 0.4$  &   \\
        \hline
    \end{tabular}
	\begin{flushright}
     	Continued on next page...
    \end{flushright}
    \begin{list}{}{}\itemsep1pt \parskip0pt \parsep0pt \footnotesize
        \item[$^{\mathrm{a}}$] Nomenclature for cluster members adopted by \citet{Lang:2005aa}.
        \item[$^{\mathrm{b}}$] Stellar identification as listed in \citet{Clark:2018aa} according to \citet{Figer:1999ab} and \citet{Liermann:2009aa}.
        \item[$^{\mathrm{c}}$] Units of right ascension are hours, minutes, and seconds, and units of declination are degrees, arcminutes, and arcseconds. Errors are in seconds and in arcseconds, respectively.
        \item[$^{\mathrm{d}}$] Flux densities measured from images corrected for primary beam attenuation.
        \item[$^{\mathrm{e}}$] Upper limit of the undetected source was fixed as 2 times the off-source rms noise level.
        \item[$^{\mathrm{f}}$] Spectral classification by \citet{Clark:2018aa}.
    \end{list}
    \label{table:quintuplet_results}

\end{sidewaystable*}

\captionsetup[table]{list=no} 
\begin{sidewaystable*}[tph!]
    \caption*{Table 3: ~--~continued from previous page}
    \footnotesize
    \setlength{\tabcolsep}{3.1pt} 
    \begin{tabular}{l l l c c c c r l}
	    \hline \hline
		\multicolumn{1}{ c }{Source} & \multicolumn{1}{ c }{Near-IR} & \multicolumn{2} { c } {Positions in X band (J2000.0) \footnotesize{$^{\mathrm{c}}$}} & \multicolumn{3} { c } {Flux density (mJy) \footnotesize{$^{\mathrm{d}}$}} & \multicolumn{1}{ c }{$\alpha$} & \multicolumn{1}{ c }{Spectral} \\ \cmidrule(lr{.5em}){3-4}  \cmidrule(lr{.5em}){5-7}
		\multicolumn{1}{ c }{name}   & \multicolumn{1}{ c }{counterpart \footnotesize{$^{\mathrm{b}}$}}  & \multicolumn{1}{ c }{R.A.} & Dec. & X-Band (2016) & X-Band (2018) & C-band (2018) & & \multicolumn{1}{ c }{type \footnotesize{$^{\mathrm{f}}$}} \\
		\hline
		qG20                                        &  LHO100           & 17 46 15.18 $\pm$ 0.02  & $-$28 49 31.39 $\pm$ 0.02  & $ 0.060 \pm 0.007$ & $0.023 \pm 0.006$ &  $0.04  \pm 0.01$  &   $-1.1 \pm 0.7$  & B2-3Ia$^{+}$   \\
        qG21                                        &  qF309            & 17 46 17.50 $\pm$ 0.02  & $-$28 49 18.61 $\pm$ 0.02  & $ 0.050 \pm 0.007$ & $0.050 \pm 0.008$ &  $0.04  \pm 0.01$  &    $0.4 \pm 0.6$  & WC8-9(d? +OB?)  \\ 
        qG22                                        &  qF274            & 17 46 17.53 $\pm$ 0.02  & $-$28 49 28.87 $\pm$ 0.02  & $ 0.050 \pm 0.007$ & $0.050 \pm 0.008$ &  $0.04  \pm 0.01$  &    $0.4 \pm 0.6$  & WN8-9ha  \\  
        qG23                                        &  qF344            & 17 46 16.67 $\pm$ 0.02  & $-$28 49 09.24 $\pm$ 0.02  & $ 0.050 \pm 0.007$ & $0.060 \pm 0.008$ &  $0.04  \pm 0.01$  &    $0.8 \pm 0.6$  & O7-8 Ia   \\
        qG24                                        &  qF278, LHO077    & 17 46 15.12 $\pm$ 0.02  & $-$28 49 34.75 $\pm$ 0.02  & $ 0.050 \pm 0.006$ & $0.060 \pm 0.009$ &  $0.03  \pm 0.01$  &    $1.4 \pm 0.7$  & B0-1 Ia$^{+}$ \\
        qG25                                        &  LHO076           & 17 46 14.14 $\pm$ 0.02  & $-$28 49 35.26 $\pm$ 0.02  & $ 0.050 \pm 0.006$ & $0.037 \pm 0.007$ &  $0.030 \pm 0.008$ &    $0.4 \pm 0.6$  & WC9d(+OB) \\
        qG26                                        &  WR 102ca         & 17 46 13.05 $\pm$ 0.02  & $-$28 49 25.10 $\pm$ 0.02  & $ 0.044 \pm 0.006$ & $0.040 \pm 0.007$ &  $0.030 \pm 0.008$ &    $0.6 \pm 0.6$  &   \\
        qG27                                        &  qF243, LHO075    & 17 46 14.13 $\pm$ 0.03  & $-$28 49 36.60 $\pm$ 0.03  & $ 0.032 \pm 0.005$ & $0.030 \pm 0.007$ &  $0.020 \pm 0.008$ &    $0.8 \pm 0.9$  & WC9d(+OB) \\
        qG28                                        &                   & 17 46 13.09 $\pm$ 0.03  & $-$28 49 29.42 $\pm$ 0.03  & $ 0.031 \pm 0.005$ & $0.030 \pm 0.007$ &  $< 0.0132$\footnotesize{$^{\mathrm{e}}$}    & $> 1.6$  &   \\
        qG29                                        &                   & 17 46 09.69 $\pm$ 0.02  & $-$28 49 39.36 $\pm$ 0.02  & $ 0.080 \pm 0.009$ & $0.17  \pm 0.02$  &  $0.15  \pm 0.02$  &    $0.2 \pm 0.3$  &    \\
        \hline
	\end{tabular}
    \begin{list}{}{}\itemsep1pt \parskip0pt \parsep0pt \footnotesize
        \item[$^{\mathrm{a}}$] Nomenclature for cluster members adopted by \citet{Lang:2005aa}.
        \item[$^{\mathrm{b}}$] Stellar identification as listed in \citet{Clark:2018aa} according to \citet{Figer:1999ab} and \citet{Liermann:2009aa}.
        \item[$^{\mathrm{c}}$] Units of right ascension are hours, minutes, and seconds, and units of declination are degrees, arcminutes, and arcseconds. Errors are in seconds and in arcseconds, respectively.
        \item[$^{\mathrm{d}}$] Flux densities measured from images corrected for primary beam attenuation.
        \item[$^{\mathrm{e}}$] Upper limit of the undetected source was fixed as 2 times the off-source rms noise level.
        \item[$^{\mathrm{f}}$] Spectral classification by \citet{Clark:2018aa}.
    \end{list}

\end{sidewaystable*} 



\subsection{Mass-loss rates}
We determined the mass-loss rates for the detected radio stars following the recipe described in \citet{Gallego-Calvente:2021yl}. We derived mass-loss rates corresponding to the observed flux densities at 6 (C-Band) and 10\,GHz (X-Band) assuming that the observed radio emission is due to free-free emission from ionised extended envelopes with a steady and completely ionised wind, with a volume filling factor $f = 1$, and an electron density profile $n_{\textrm{e}}$ $\propto$ $r^{-2}$. In the case of non-thermal contributions our values, showed in Table\,\ref{table:quintuplet_mass-loss}, represent upper limits to the true mass-loss rates.

For most Quintuplet cluster members, we can assume that helium stays singly ionised in the radio emitting region of the stellar wind, so the number of free electrons per ion and the mean ionic charge can be set $\gamma = 1 = Z$ \citep[][]{Leitherer:1997yf}. However, in the case of Pistol Star and qG3 (the two luminous blue variables (LBVs)) accounting for the fact that helium is predominantly neutral in the winds of these cool stars, we take $\gamma = 0.8$, and $Z = 0.9$ similarly to other LBVs studies \citep[e.g.][]{Leitherer:1995aa, Agliozzo:2019aa}.

\begin{sidewaystable*}[tph!]
	\caption[Quintuplet cluster: Radio mass-loss rates]{Radio mass-loss rates in the C-band (2018) and in the two X-band epochs (2016/2018).}	
	\footnotesize
	\begin{tabular}{llcccccc}
	 \hline \hline
	 \noalign{\smallskip}
         Source  & Spectral                                 & $v_{\infty}$ \footnotesize{$^{\mathrm{c}}$}   & $\mu$ \footnotesize{$^{\mathrm{c}}$}  & $\dot{M}_{2018}$  & $\dot{M}_{2016}$ & $\dot{M}_{2018}$ &  $\dot{M}_{\textrm{Lang\,2005}}$            \\
        \noalign{\smallskip}
                 & type \footnotesize{$^{\mathrm{b}}$}     & (Km/s)                                         & (\#)                                  & 6.0\,GHz          & 10.0\,GHz        & 10.0\,GHz        &  22.5\,GHz \footnotesize{$^{\mathrm{d}}$}   \\
     \noalign{\smallskip}
	 \hline
     \noalign{\smallskip}
        qG1 (QR6 \footnotesize{$^{\mathrm{a}}$})        & B1-2 Ia$^{+}$         &  500      & 1.3   & ($2.0 \pm 0.4) \times 10^{-5}$     & ($3.8 \pm 0.8) \times 10^{-5}$ & ($2.1 \pm 0.5) \times 10^{-5}$  & $1.2 \times 10^{-5}$   \\
        qG2 (QR5 \footnotesize{$^{\mathrm{a}}$})        & WN11h                 &  300      & 1.8   & ($1.3 \pm 0.3) \times 10^{-5}$     & ($1.8 \pm 0.4) \times 10^{-5}$ & ($1.8 \pm 0.4) \times 10^{-5}$  & $3.0 \times 10^{-5}$   \\ 
        qG3 (QR9 \footnotesize{$^{\mathrm{a}}$})        & LBV                   &  170      & 1.9   & ($5.9 \pm 1.3) \times 10^{-6}$     & ($1.1 \pm 0.2) \times 10^{-5}$ & ($1.3 \pm 0.3) \times 10^{-5}$  & $5.2 \times 10^{-6}$   \\
        Pistol Star                                     & LBV                   &  105      & 2.2   & ($2.8 \pm 0.6) \times 10^{-6}$     & ($5.8 \pm 1.0) \times 10^{-6}$ & ($7.1 \pm 1.0) \times 10^{-6}$  & $3.1 \times 10^{-5}$   \\
        qG4                                             & WN10h                 &  400      & 1.8   & ($9.2 \pm 2.0) \times 10^{-6}$     & ($1.3 \pm 0.3) \times 10^{-5}$ & ($1.2 \pm 0.3) \times 10^{-5}$  &                        \\
        qG5                                             & <B0 I                 &  500      & 1.3   & ($1.5 \pm 0.3) \times 10^{-5}$     & ($1.2 \pm 0.3) \times 10^{-5}$ & ($1.0 \pm 0.2) \times 10^{-5}$  &                        \\
        qG6 (QR8 \footnotesize{$^{\mathrm{a}}$})        & WN9h                  &  700      & 2.78  & ($2.7 \pm 0.6) \times 10^{-5}$     & ($3.5 \pm 0.7) \times 10^{-5}$ & ($3.5 \pm 0.7) \times 10^{-5}$  & $4.2 \times 10^{-5}$   \\
        qG7                                             & O7-8 Ia$^{+}$/WNLh    &  1500     & 1.3   & ($3.3 \pm 0.7) \times 10^{-5}$     & ($2.9 \pm 0.6) \times 10^{-5}$ & ($3.0 \pm 0.7) \times 10^{-5}$  &                        \\
        qG8                                             & WC8-9(d? + OB?)       &  1250     & 4.7   & ($9.8 \pm 2.0) \times 10^{-5}$     & ($7.9 \pm 2.0) \times 10^{-5}$ & ($1.1 \pm 0.2) \times 10^{-4}$  &                        \\
        qG9                                             & WC9d (+OB)            &  1250     & 4.7   & ($5.4 \pm 1.0) \times 10^{-5}$     & ($6.7 \pm 1.0) \times 10^{-5}$ & ($5.8 \pm 1.0) \times 10^{-5}$  &                        \\
        qG10                                            & WN8-9ha               &  900      & 2.0   & ($1.2 \pm 0.3) \times 10^{-5}$     & ($2.0 \pm 0.4) \times 10^{-5}$ & ($1.7 \pm 0.4) \times 10^{-5}$  &                        \\
        qG11 (QR7 \footnotesize{$^{\mathrm{a}}$})       & WC9d(+OB)             &  1250     & 4.7   & ($5.4 \pm 1.0) \times 10^{-5}$     & ($5.8 \pm 1.0) \times 10^{-5}$ & ($4.5 \pm 1.0) \times 10^{-5}$  & $2.8 \times 10^{-4}$   \\
        qG12                                            & O7-8 Ia               &  1500     & 1.3   & ($1.5 \pm 0.3) \times 10^{-5}$     & ($1.9 \pm 0.4) \times 10^{-5}$ & ($2.3 \pm 0.5) \times 10^{-5}$  &                        \\
        qG13 (QR4 \footnotesize{$^{\mathrm{a}}$})       & B1-2 Ia$^{+}$/WNLh    &  300      & 1.3   & ($2.6 \pm 0.6) \times 10^{-6}$     & ($3.6 \pm 0.8) \times 10^{-6}$ & ($3.3 \pm 0.7) \times 10^{-6}$  & $2.9 \times 10^{-5}$   \\
        qG14                                            & B1-2 Ia$^{+}$         &  500      & 1.3   & ($4.9 \pm 1.0) \times 10^{-6}$     & ($6.0 \pm 1.0) \times 10^{-6}$ & ($6.0 \pm 1.0) \times 10^{-6}$  &                        \\
        qG15                                            & WC8(d? +OB?)          &  1250     & 4.7   & ($3.5 \pm 0.7) \times 10^{-5}$     & ($4.9 \pm 1.0) \times 10^{-5}$ & ($4.9 \pm 1.0) \times 10^{-5}$  &                        \\
        \noalign{\smallskip}
        \hline
	\end{tabular}
	\begin{flushright}
     	Continued on next page...
    \end{flushright}
    \begin{list}{}{}\itemsep1pt \parskip0pt \parsep0pt \footnotesize
        \item[\textbf{Notes.}] It is shown a comparison to estimations done by \citep{Lang:2005aa} from their 22.5\,GHz and their 8.5\,GHz observations, re-scaled for taking into account their assumed $v_{\infty}$ = 1000\,Km\,s$^{-1}$ and $\mu$ = 2. All mass-loss rates, $\dot{M}$, are in units of M$_{\odot}$\,yr$^{-1}$.
        \item[$^{\mathrm{a}}$] Nomenclature for cluster members adopted by \citet{Lang:2005aa}.
        \item[$^{\mathrm{b}}$] Spectral classification by \citet{Clark:2018aa}.
        \item[$^{\mathrm{c}}$] Wind parameters for qG3 and Pistol stars from Table\,1 by \citet{Najarro:2009aa}; values for qG8, qG9, qG11, qG15, qG16, qG21, qG25, and qG27 based on measurements for GSC4 star by \citet{Najarro:2017aa}; parameters for qG2, qG4, qG6, qG10, and qG22 by \citet{Liermann:2010aa}, except $\mu$ for qG2 and qG4 estimated by Najarro (private communication); for sources of spectral types WN6, B supergiants and OIa we adopted values of stars of similar type.
        \item[$^{\mathrm{d}}$] Derived from the 22.5\,GHz flux density for qG1 (QR6), qG2 (QR5), qG6 (QR8) sources and Pistol Star, and from the 8.5\,GHz flux density for qG13 (QR4), qG11 (QR7) and qG3 (QR9), as described in Table\,4 by \citet{Lang:2005aa}. Values have been re-scaled (see above).
    \end{list}
	\label{table:quintuplet_mass-loss}
\end{sidewaystable*}

\captionsetup[table]{list=no} 
\begin{sidewaystable*}[tph!]
	\caption*{Table 4.4: ~--~continued from previous page}	
	\footnotesize
	\begin{tabular}{llcccccc}
	 \hline \hline
	 \noalign{\smallskip}
         Source  & Spectral                                 & $v_{\infty}$ \footnotesize{$^{\mathrm{c}}$}   & $\mu$ \footnotesize{$^{\mathrm{c}}$}  & $\dot{M}_{2018}$  & $\dot{M}_{2016}$ & $\dot{M}_{2018}$ &  $\dot{M}_{\textrm{Lang\,2005}}$            \\
        \noalign{\smallskip}
                 & type \footnotesize{$^{\mathrm{b}}$}     & (Km/s)                                         & (\#)                                  & 6.0\,GHz          & 10.0\,GHz        & 10.0\,GHz        &  22.5\,GHz \footnotesize{$^{\mathrm{d}}$}   \\
     \noalign{\smallskip}
	 \hline
     \noalign{\smallskip}
		qG16                                            & WC8(d? +OB?)          &  1250     & 4.7   & ($5.4 \pm 1.0) \times 10^{-5}$     & ($4.9 \pm 1.0) \times 10^{-5}$ & ($5.8 \pm 1.0) \times 10^{-5}$  &                        \\
        qG17                                            & B0-1 Ia$^{+}$/WNLh    &  500      & 1.3   & ($3.2 \pm 0.8) \times 10^{-6}$     & ($5.1 \pm 1.0) \times 10^{-6}$ & ($4.8 \pm 2.0) \times 10^{-6}$  &                        \\
        qG18                                            & WN6                   &  1600     & 4.1   & ($2.6 \pm 0.7) \times 10^{-5}$     & ($5.1 \pm 1.0) \times 10^{-5}$ & ($4.8 \pm 2.0) \times 10^{-5}$  &                        \\
        qG19                                            &                       &           &       &                                    &                                &                                 &                        \\
        qG20                                            & B2-3Ia$^{+}$          &  500      & 1.3   & ($3.2 \pm 0.9) \times 10^{-6}$     & ($4.4 \pm 1.0) \times 10^{-6}$ & ($2.1 \pm 0.6) \times 10^{-6}$  &                        \\
        qG21                                            & WC8-9(d? +OB?)        &  1250     & 4.7   & ($2.9 \pm 0.8) \times 10^{-5}$     & ($3.5 \pm 0.8) \times 10^{-5}$ & ($3.5 \pm 0.8) \times 10^{-5}$  &                        \\
        qG22                                            & WN8-9ha               &  900      & 2.0   & ($9.0 \pm 2.0) \times 10^{-6}$     & ($1.1 \pm 0.2) \times 10^{-5}$ & ($1.1 \pm 0.3) \times 10^{-5}$  &                        \\
        qG23                                            & O7-8 Ia               &  1500     & 1.3   & ($9.7 \pm 3.0) \times 10^{-6}$     & ($1.2 \pm 0.3) \times 10^{-5}$ & ($1.3 \pm 0.3) \times 10^{-5}$  &                        \\
        qG24                                            & B0-1 Ia$^{+}$         &  500      & 1.3   & ($2.6 \pm 0.8) \times 10^{-6}$     & ($3.8 \pm 0.9) \times 10^{-6}$ & ($4.4 \pm 1.0) \times 10^{-6}$  &                        \\
        qG25                                            & WC9d(+OB)             &  1250     & 4.7   & ($2.4 \pm 0.7) \times 10^{-5}$     & ($3.2 \pm 0.7) \times 10^{-5}$ & ($2.8 \pm 0.7) \times 10^{-5}$  &                        \\
        qG26                                            &                       &           &       &                                    &                                &                                 &                        \\
        qG27                                            & WC9d(+OB)             &  1250     & 4.7   & ($1.7 \pm 0.6) \times 10^{-5}$     & ($2.5 \pm 0.6) \times 10^{-5}$ & ($2.4 \pm 0.6) \times 10^{-5}$  &                        \\
        qG28                                            &                       &           &       &                                    &                                &                                 &                        \\
        qG29                                            &                       &           &       &                                    &                                &                                 &                        \\
        \noalign{\smallskip}
        \hline
	\end{tabular}
    \begin{list}{}{}\itemsep1pt \parskip0pt \parsep0pt \footnotesize
        \item[\textbf{Notes.}] It is shown a comparison to estimations done by \citep{Lang:2005aa} from their 22.5\,GHz when possible, and otherwise from their 8.5\,GHz observations, re-scaled for taking into account their assumed $v_{\infty}$ = 1000\,Km\,s$^{-1}$ and $\mu$ = 2. All mass-loss rates, $\dot{M}$, are in units of M$_{\odot}$\,yr$^{-1}$.
        \item[$^{\mathrm{a}}$] Nomenclature for cluster members adopted by \citet{Lang:2005aa}.
        \item[$^{\mathrm{b}}$] Spectral classification by \citet{Clark:2018aa}.
        \item[$^{\mathrm{c}}$] Wind parameters for qG3 and Pistol stars from Table\,1 by \citet{Najarro:2009aa}; values for qG8, qG9, qG11, qG15, qG16, qG21, qG25, and qG27 based on measurements for GSC4 star by \citet{Najarro:2017aa}; parameters for qG2, qG4, qG6, qG10, and qG22 by \citet{Liermann:2010aa}, except $\mu$ for qG2 and qG4 estimated by Najarro (private communication); for sources of spectral types WN6, B supergiants and OIa we adopted values of stars of similar type.
        \item[$^{\mathrm{d}}$] Derived from the 22.5\,GHz flux density for qG1 (QR6), qG2 (QR5), qG6 (QR8) sources and Pistol Star, and from the 8.5\,GHz flux density for qG13 (QR4), qG11 (QR7) and qG3 (QR9), as described in Table\,4 by \citet{Lang:2005aa}. Values have been re-scaled (see above).
    \end{list}
\end{sidewaystable*}

Table\,\ref{table:quintuplet_mass-loss} also shows the estimations done by \cite{Lang:2005aa} from their 22.5\,GHz flux densities when possible, and otherwise from their 8.5\,GHz flux densities. They used their higher frequency observations as better tracer of the thermal component and so give more reliable mass-loss rates because contamination of the flux density by the non-thermal component can occur at the lower frequencies (according to \citealt{Contreras:1996aa}). \citet{Lang:2005aa} assumed a terminal wind velocity of 1000\,Km\,s$^{-1}$ and a mean molecular weight equal to 2 for all sources. Thus, in order to compare our data with their data, we have re-scaled Lang's values multiplying them by $v_{\infty}$/1000 and $\mu$/2, where $v_{\infty}$ and $\mu$ are the values adopted in this work. We derived a significantly lower mass-loss rate for the Pistol star than \citet{Lang:2005aa} because the latter work reports a roughly ten times higher flux density for this source. This may be explained through variability of the source (see below) or,  possibly also, the lower angular resolution of the observations of \citet{Lang:2005aa} which may have resulted in the flux of the Pistol star being contaminated by flux coming from the Pistol nebula, where it is embedded. Similar factor $\sim$\,10 discrepancies exist for the sources qG13/QR4 and qG11/QR7.



\section{Properties of the sources \label{sec4:properties}}
According to the spectral classification, most of detected sources in the Quintuplet cluster are young, massive, post-main sequence stars, being the majority of the Wolf-Rayet spectral type (see Table\,\ref{table:quintuplet_results}). In particular, two luminous blue variable, eight WC, six WN, seven B supergiants and three sources of type O\,Ia have been found.


After considering the inferred spectral indices, even though the corresponding uncertainties are relatively high, we show that qG1, qG2, qG4, qG6, qG9, qG10, qG12$-$15, qG17, qG18, and qG21$-$27 have spectral indices consistent with thermal emission from ionised stellar winds. The divergence of $\alpha$ from the canonical value of 0.6 can occur for a number of reasons, such as deviations in the wind conditions possibly due to the presence of condensations (\textit{clumps}) that produce a non-standard electron density profile ($n_{\textrm{e}}$ $\propto$ $r^{-s}$, with $s\,\neq\,2$) and/or changes in the run of ionisation or wind geometry with radius due to internal shocks. \citet{Leitherer:1997yf} suggested that there is little variation in the mass-loss rates for Wolf-Rayet stars, with an averaged value of $\sim$\,4\,$\times$\,10$^{-5}$\,M$_{\odot}$\,yr$^{-1}$. The WR stars from our data set have $\dot{M}$ values ranging from 1.3\,$\times$\,10$^{-5}$ to 7.9\,$\times$\,10$^{-5}$ M$_{\odot}$\,yr$^{-1}$, which agrees well with the results of \citet{Leitherer:1997yf}, as well as for other young, massive stellar types. This conclusion lead us to choose B1$-$2 Ia$^{+}$ and B0$-$1 Ia$^{+}$ as the most probable spectral types for qG13 and qG17, respectively, as their mass-loss rates are not in that range.

On the other hand, qG5, qG7, qG11, qG19, and qG20 have flat or inverted spectral indices, which indicates the presence of non-thermal emission that may be attributed to colliding winds in binaries. 

qG8, and qG16 are ambiguous cases. No measurement of $\alpha$ is available for the star qG28 because it is only detected in a single band. 

We can probe the existence of long-term variability since we have two epochs with X-band measurements. We contemplate the existence of flux density variability when the differences in the flux densities between the two epochs are higher than 5\,$\sigma$. With this criterion we consider the two LBVs, qG3 and the Pistol star as variable sources, as well as the  sources qG1, qG8, qG12, qG20, and qG29. For qG10, and qG11, the variability is rather small, but taking into account the possible binary nature of qG11, we speculate that the contribution from non-thermal emission may be modulated by the orbital motion of the system. Further investigation would be necessary to find a correlation between the periodicity and the orbital period. In conclusion, we find that 9 out of the 29 radio sources, approximately 30\%, display significant variability. This is a remarkably higher fraction than in the Arches cluster, where we find that $\lesssim$\,15\% of the radio stars display variability \citep[see][]{Gallego-Calvente:2021yl}. We speculate that this may be related to the advanced evolutionary state of the older Quintuplet stars.


\section{Analysis of the radio luminosity function \label{sec:RLF}}

\subsection{Basic assumptions}

With assumptions about the age and mass of a young cluster, a comparison between  the number and the observed flux of detected radio stars with predictions from theoretical stellar evolutionary models (isochrones) can help to assess the quality of those models, and also, if the theoretical isochrones are considered sufficiently reliable, to infer basic properties of the cluster. In \citet{Gallego-Calvente:2021yl} we have shown that the observed number of radio stars in the Arches cluster appears to require a top-heavy initial mass function with a power-law exponent $\alpha_{IMF}=-1.8$ (at odds with the Solar environment ``Salpeter'' exponent $\alpha_{IMF}=-2.35$). 

In this work, we tentatively take a step further and use the radio luminosity function (RLF) of the stars, i.e.\ we use both the number of sources as well as their flux density for comparison with the models. We have done this first for the Arches cluster and then for the Quintuplet cluster. We emphasise that we cannot undertake an exhaustive study here, which would require taking into account a large number of potentially important parameters, such as, for example, the role of metallicity or the wind volume filling factor, which we neglect here. Nevertheless, we show that our simple toy modelling, with these basic assumptions, indicates that current theoretical models of the evolution of massive stars appear to be fairly realistic, and that we can infer  constraints on the properties of the clusters by comparing radio observations with predictions from isochrones.

 We use PARSEC\footnote{http://stev.oapd.inaf.it/cmd} \citep[release v$1.2$S $+$ COLIBRI S$\_37$,][]{Bressan:2012xy,Chen:2014nr,Chen:2015sf,Tang:2014rm,Marigo:2017fb,Pastorelli:2019er,Pastorelli:2020pt} and MIST\footnote{https://waps.cfa.harvard.edu/MIST/} \citep[][]{Dotter:2016sh,Choi:2016qh,Paxton:2011vh,Paxton:2013vb,Paxton:2015hz} theoretical isochrones for solar metallicity. To convert the mass-loss rates from the theoretical isochrones to observable radio flux, we use the relation
 
\begin{eqnarray}
\label{equ:S_nu}
    \lefteqn{\biggl[ \frac{S_{\nu}}{mJy}
    \biggr] = (5.34 \times 10^{-4})^{-4/3}\,
    f_{clump}^{1/2}\,\biggl[\frac{\dot{M}}{M_{\odot}\,yr^{-1}} \biggr] 
    \biggl[ \frac{v_{\infty}}{km\,s^{-1}}\biggr]^{-4/3}}
    \nonumber\\
    & & {} \hspace{0.8cm} \biggl[ \frac{d}{kpc}\biggr]^{-2}\,
    \biggl[ \frac{\nu}{Hz}\biggr]^{2/3}\,\biggl[ \frac{{\mu}^{2}}{Z^{2} \gamma g_{\nu}}\biggr]^{-2/3},
\end{eqnarray}
\vspace{0.2cm}

\noindent which has been reported before in similar versions \citep[e.g.][]{Gallego-Calvente:2021yl,Montes:2009aa,Leitherer:1997yf}. The observing frequency is $\nu=10$\,GHz. We assume typical values, with a  mean molecular weight $\mu=1.3$, a mean ion charge $Z=1$, and a mean number of electrons per ion $\gamma=1$. As concerns the Gaunt factor
\begin{equation}
g_{\nu} = 9.77 \cdot \displaystyle\biggl(1 + 0.13 \cdot log \,\frac{T_e^{3/2}}{Z \nu}\displaystyle\biggr),
\end{equation}

\noindent we use $T_{e}=10^{4}$\,K. Any changes to these assumed values of the parameters by factors of two to a few do not significantly change the resulting radio flux. A potentially important factor, that we have not yet mentioned, is wind clumping. The volume filling factor, $f_{clump}$, is one for a homogeneous stellar wind and smaller than one for a clumpy wind. For simplicity, we here assume $f_{clump}=1$. 

\begin{figure}[ht!]
    \includegraphics[width=0.5\textwidth]{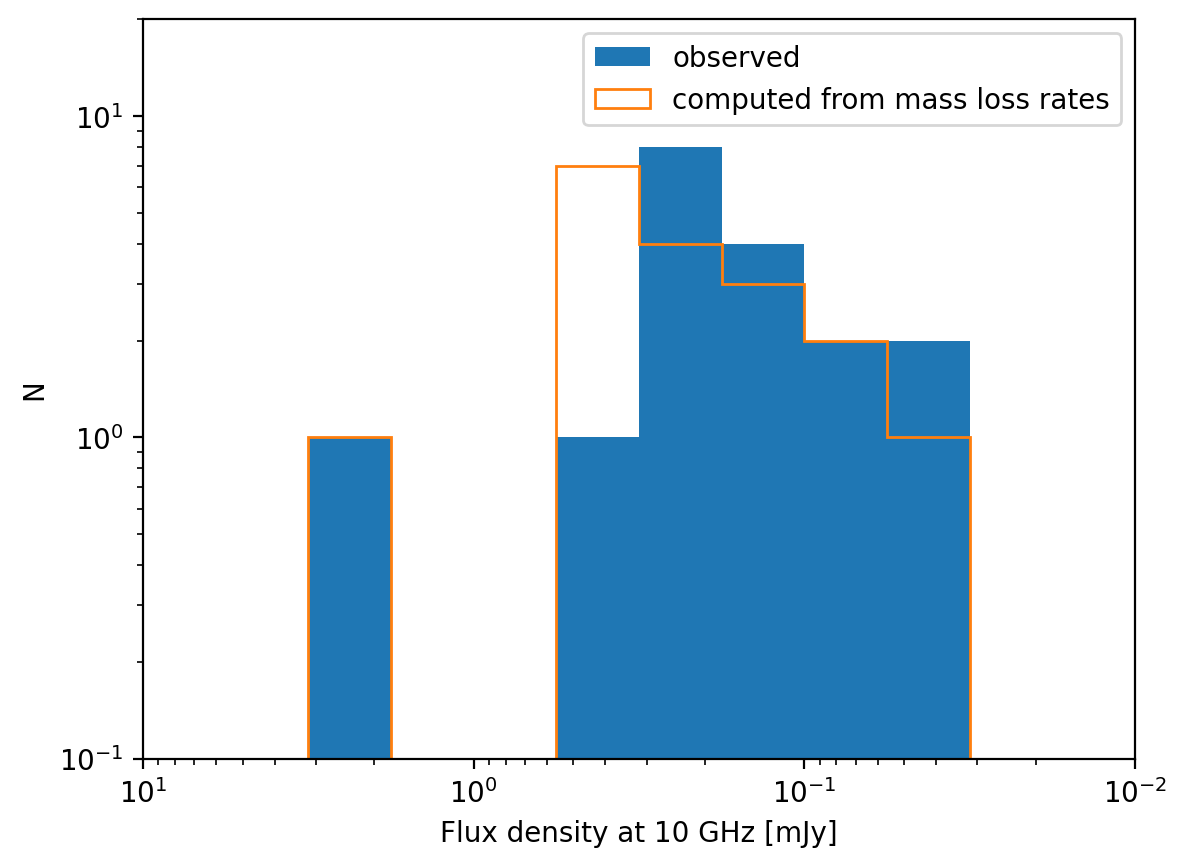}
    \caption[Histogram of S$_{\nu}$ of observed radio stars in Arches cluster]{Blue, solid: Histogram of observed flux densities of radio stars in the Arches cluster \citep{Gallego-Calvente:2021yl}. Orange, outline: Histogram of radio fluxes of the same stars, using the mass-loss rates reported by \citet{Gallego-Calvente:2021yl} and converting to radio flux density with Equation\,\ref{equ:S_nu} assuming the same parameters for all stars.}
	\label{fig:flux_obs_theor}
\end{figure}

The validity of such a rule-of-thumb approach is demonstrated in Fig.\,\ref{fig:flux_obs_theor}, where we compare the observed flux densities of radio stars in the Arches cluster with the ones computed from their individually determined mass-loss rates \citep{Gallego-Calvente:2021yl} and using the simplified assumptions. The differences appear to be small enough to justify an efficient, simplified computation of the radio flux densities. From here on we will use these simplified assumptions on the parameters that enter Equation\,\ref{equ:S_nu} and only consider the observed radio flux densities, not the mass-loss rates.

For our cluster models we always use a one-segment power-law initial mass function (IMF) with an upper mass of 150\,M$_{\odot}$ and a lower mass of 0.8\,M$_{\odot}$. Our results are not sensitive to reasonable changes of these parameters. Also, since our sources of interest represent only the very tip of the mass distribution, assuming a two-segment IMF does not introduce any significant changes.

We tested whether the random sampling of the high mass end of the IMF is biasing our results by random realisations of clusters of given age, mass, and IMF. We only used MIST models for this test and assumed that we could only detect stars with a radio flux density $\gtrsim$ 0.04\,mJy, the approximate detection limit here and in \citet{Gallego-Calvente:2021yl}. We found that the age of the clusters can typically be recovered well ($\pm$\,1\,Myr) and that the cluster parameters could best be recovered for an age of $\sim$\,3\,Myr, where the number of bright radio stars peaks.  Whether the IMF was top-heavy or not could also be recovered with great reliability. However, we found a  degeneracy between the RLFs for the ages of 2 and 6 Myrs. Since these values lie safely outside (or at the very extreme) of the ages reported for the Arches and Quintuplet clusters, we do not use them in our subsequent analyses.

\subsection{Models}

\begin{figure}[ht!]
    \includegraphics[width=0.5\textwidth]{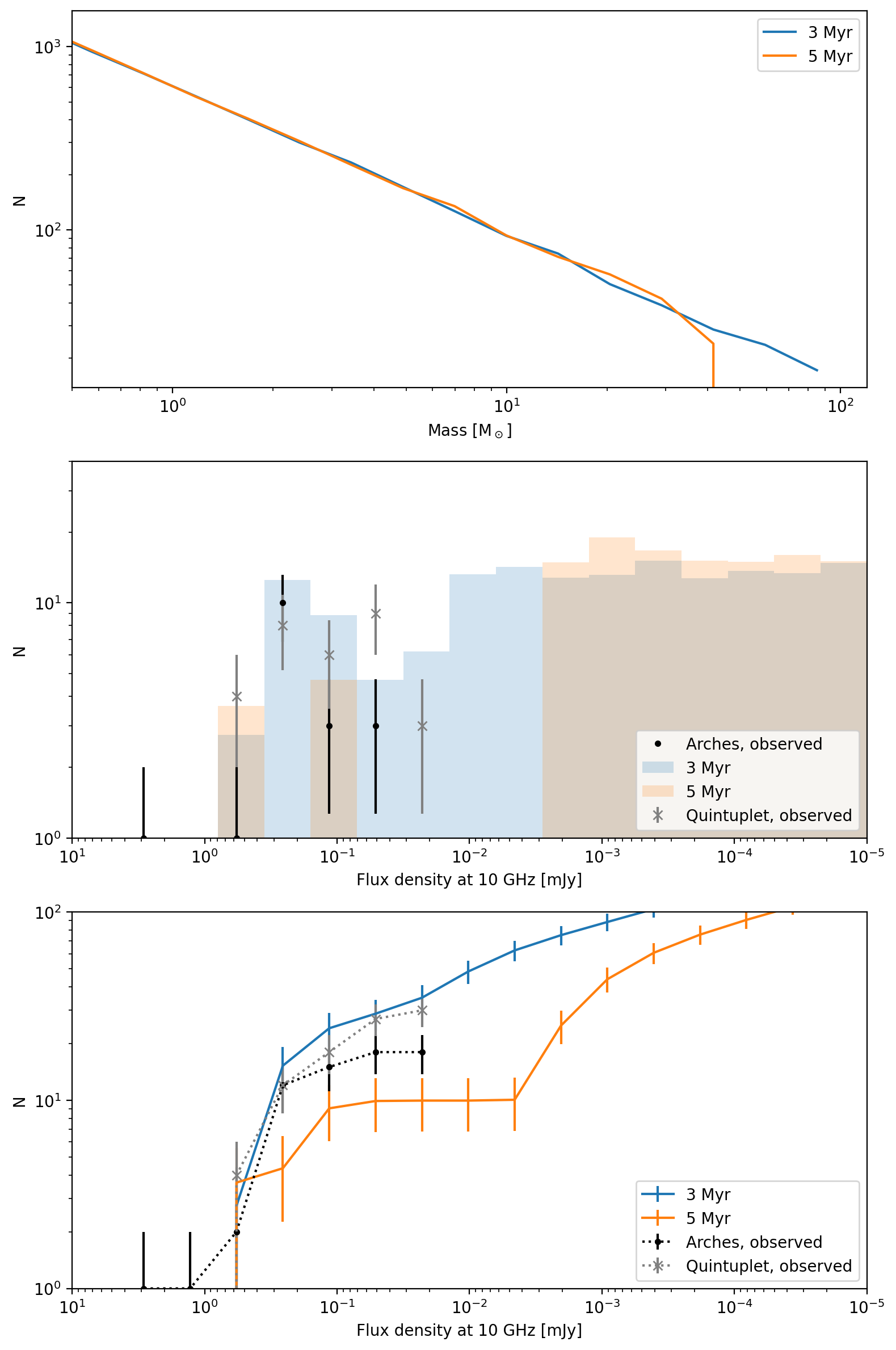}
    \caption[S$_{\nu}$ of massive stars in a model cluster, using MIST isochrones]{Radio flux densities of massive stars in a model cluster, using MIST isochrones. Top: Present-day mas function. Middle: Histograms of predicted radio flux densities. Bottom: Cumulative histograms of predicted radio flux densities. The black histogram are observed values for the Arches cluster \citep[dotted]{Gallego-Calvente:2021yl} and for the Quintuplet cluster (this work, solid).}
	\label{fig:age_radioflux}
\end{figure}

\begin{figure}[ht!]
    \includegraphics[width=0.5\textwidth]{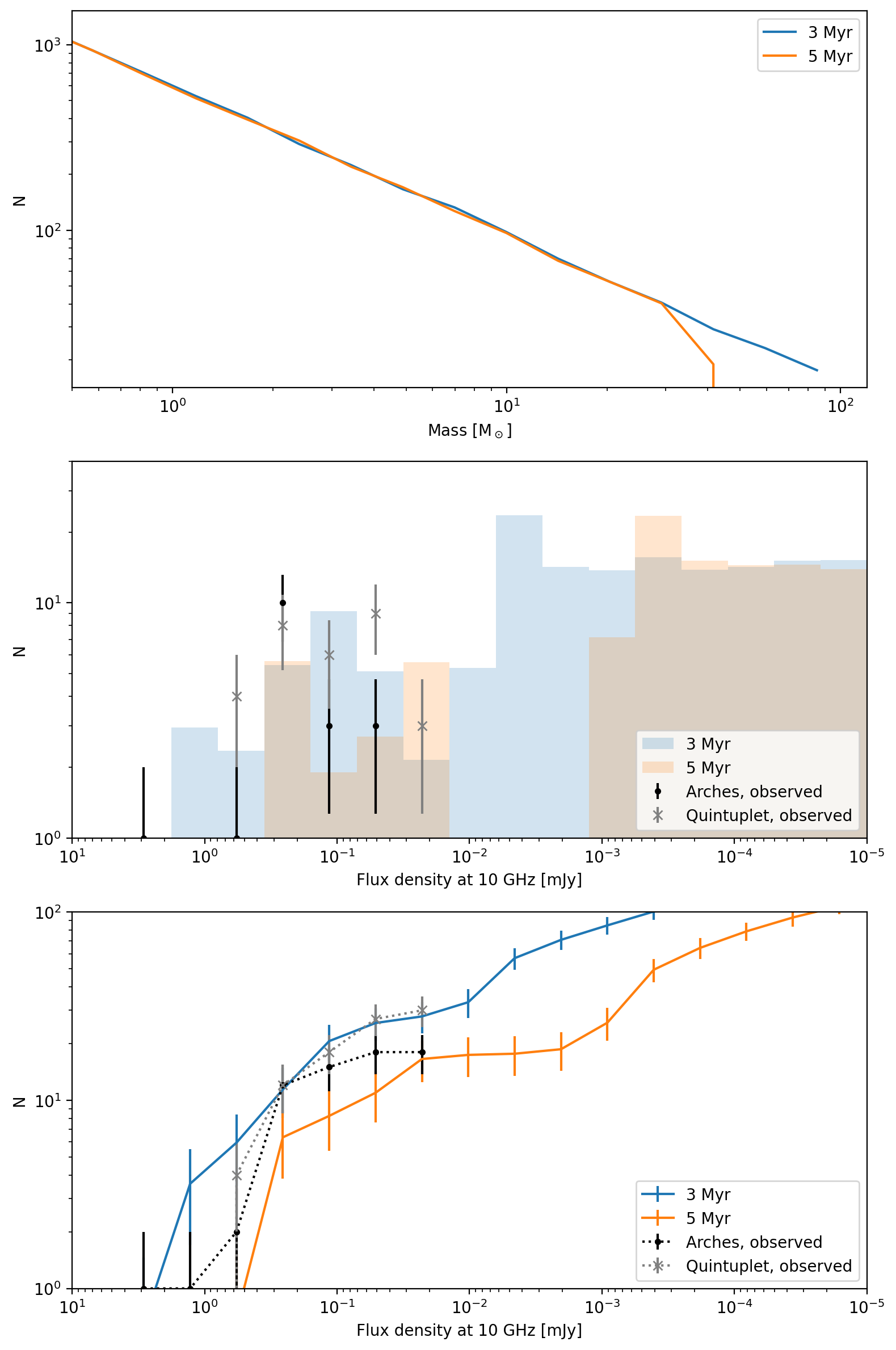}
    \caption[S$_{\nu}$ of massive stars in a model cluster, using PARSEC isochrones]{Radio flux densities of massive stars in a model cluster, using PARSEC isochrones. Top: Present-day mas function. Middle: Histograms of predicted radio flux densities. Bottom: Cumulative histograms of predicted radio flux densities. The black histogram are observed values for the Arches cluster \citep[dotted]{Gallego-Calvente:2021yl} and for the Quintuplet cluster (this work, solid).}
	\label{fig:age_radioflux_Parsec}
\end{figure}

Figure\,\ref{fig:age_radioflux} shows predictions at two different ages for a model cluster with a total initial stellar mass of  10$^{4}$\,M$_{\odot}$ and a power-law IMF with an exponent $\alpha_{IMF}= - 1.8$ \citep{Hosek:2019vn}, using MIST isochrones. We averaged and used the 1\,$\sigma$ confidence intervals of the result of 10  simulations of the cluster. The upper panel shows the present-day mass function, the middle one histograms of the  predicted radio flux density, and the lower one cumulative histograms of the predicted radio flux density. Observed values in Arches and Quintuplet are overplotted in black. Finally, we show the same models, created with PARSEC isochrones, in Figure\,\ref{fig:age_radioflux_Parsec}. We can see differences due to the use of different theoretical isochrones, but we can also see the overall similarity of the predictions.

Given the low flux density of radio stars and the sensitivity of our observations we can only observe the bright tail of the distribution.  We show that the RLF can change drastically as a function of age. This is due to the rapid development of the extremely massive stars at such young ages. The latter also means that single stellar evolution models are expected to work well because we do not expect any significant effects such as mass-transfer to occur in binaries at these time scales. Also, stellar evolution happens so fast at these young ages that the radio flux of a binary will typically be dominated by one of the two sources, except in the unlikely case that the two stellar masses are exactly equal. Of course, colliding wind binaries will be a source of uncertainty, but their numbers in our sample are small according to the reported spectral indices and we do not think that they will affect the observed population statistics significantly. We do not test this assumption here, however, given that this would go beyond the simple model presented here. To avoid having to deal with complex binning issues we will use cumulative luminosity functions from this point on.

\begin{figure}[ht!]
    \includegraphics[width=0.5\textwidth]{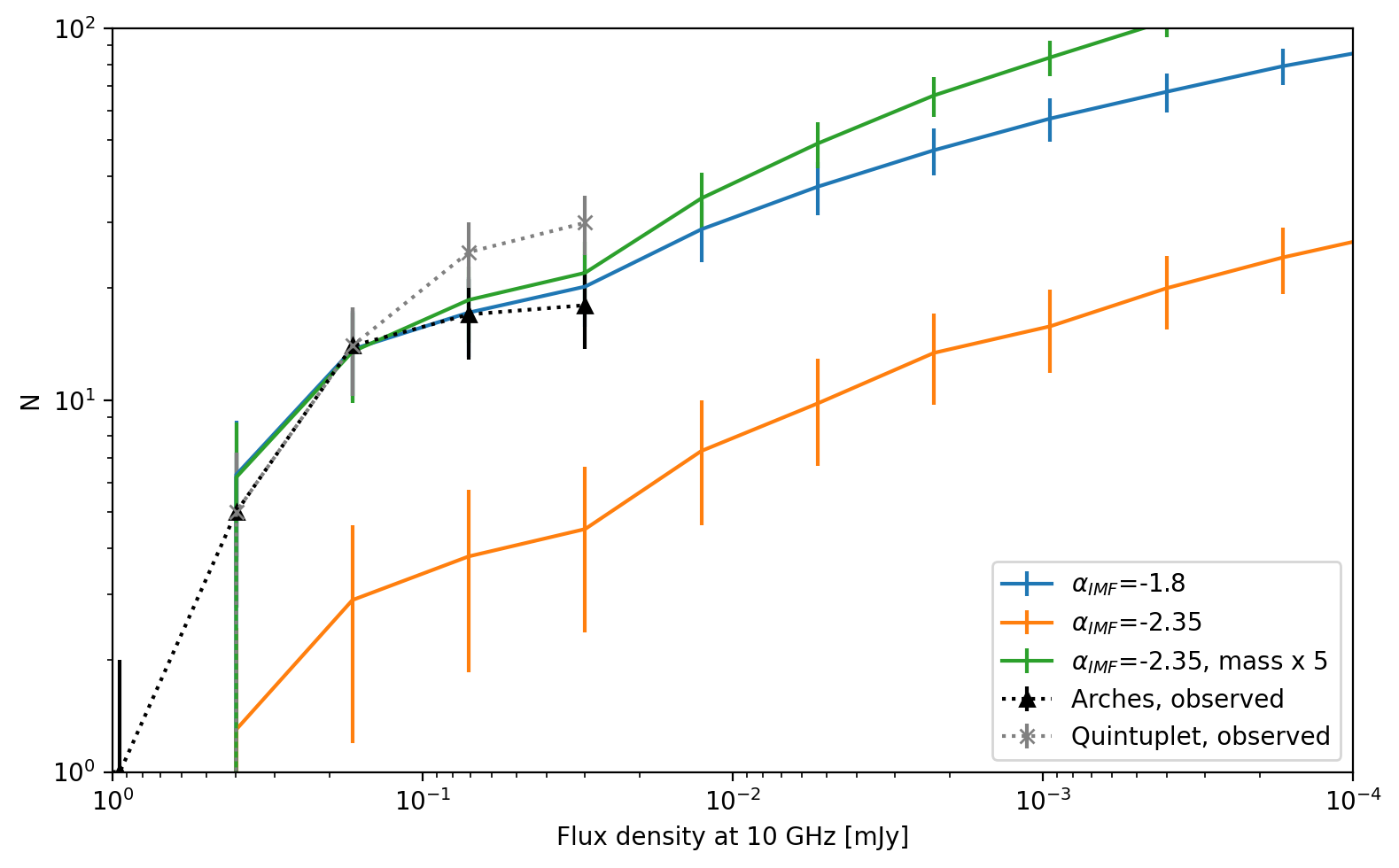}
    \caption[Simulated radio luminosity functions]{Simulated radio luminosity functions for a 3\,Myr old cluster with an initial stellar mass of $10^{4}$\,M$_{\odot}$, using $\alpha_{IMF}=-1.8$ and the standard Salpeter value of $\alpha_{IMF}=-2.35$.}
    \label{fig:radio_alphas}
\end{figure}

A parameter of significant interest is the exponent of the power-law of the IMF, $\alpha_{IMF}$, because it has a strong influence on the number of massive stars that initially form the cluster. In \cite{Gallego-Calvente:2021yl} we showed that the observed number of radio stars in the Arches cluster favoured a top-heavy IMF. In Fig.\,\ref{fig:radio_alphas} we illustrate how the bright end of the RLF changes as a function of $\alpha_{IMF}$. The RLF of the model cluster with a Salpeter IMF will only be similar to the one of a cluster with a top-heavy IMF ($\alpha_{IMF}=-1.8$) for an initial mass that is four to five times higher. While uncertainties up to a factor two are possible in the observational determination of a young, massive cluster's mass (for example from optical or infrared observations), higher factors can generally be ruled out safely. Therefore, the observed RLF can help to constrain the exponent of the power-law IMF of a young massive cluster. The most important degeneracy to keep in mind is the one between cluster mass and IMF exponent. There is also some degeneracy with age, but less relevant, because the shape of the bright end of the RLF changes sensitively as a function of age. With reasonable constraints on the age and mass of a cluster, one may therefore constrain its properties via the RLF.

Following the previous considerations, we proceed to fit cluster models to the observed radio luminosity functions of the Arches and Quintuplet clusters. This is a demonstration of the principle, so we limit ourselves to a crude exploration of parameter space with simple, brute-force Monte Carlo simulations. To create the MC samples, the observed radio flux densities were varied randomly assuming for each source a Gaussian normal distribution centred on the observed radio flux density and a conservative standard deviation of 30\% from the mean value (assuming the same uncertainty for all sources). The parameter space was spanned by two values of $\alpha_{IMF}=-1.8, -2.35$, ages of 2.5, 3, 3.5, 4, 4.5, 5\,Myr and cluster masses of 1.0 and 1.5\,$\times$\,10$^{4}$\,M$_{\odot}$.

\subsection{Arches cluster}

\begin{figure}[ht!]
    \includegraphics[width=0.5\textwidth]{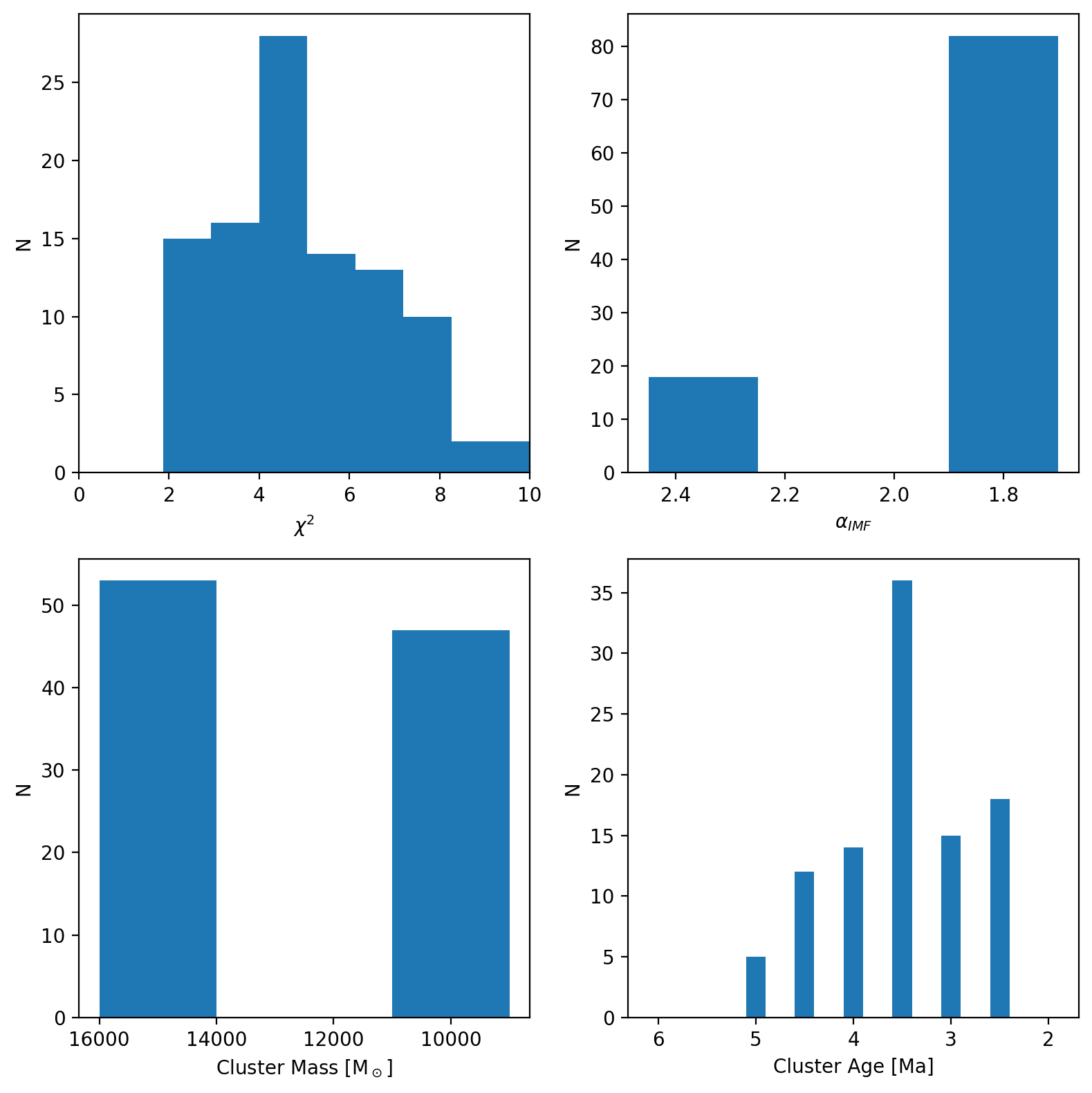}
    \caption[MC simulations of the Arches RLF, using MIST isochrones]{Results of MC simulations of the Arches RLF, using MIST isochrones.}
    \label{fig:Arches_MIST}
\end{figure}

\begin{figure}[ht!]
    \includegraphics[width=0.5\textwidth]{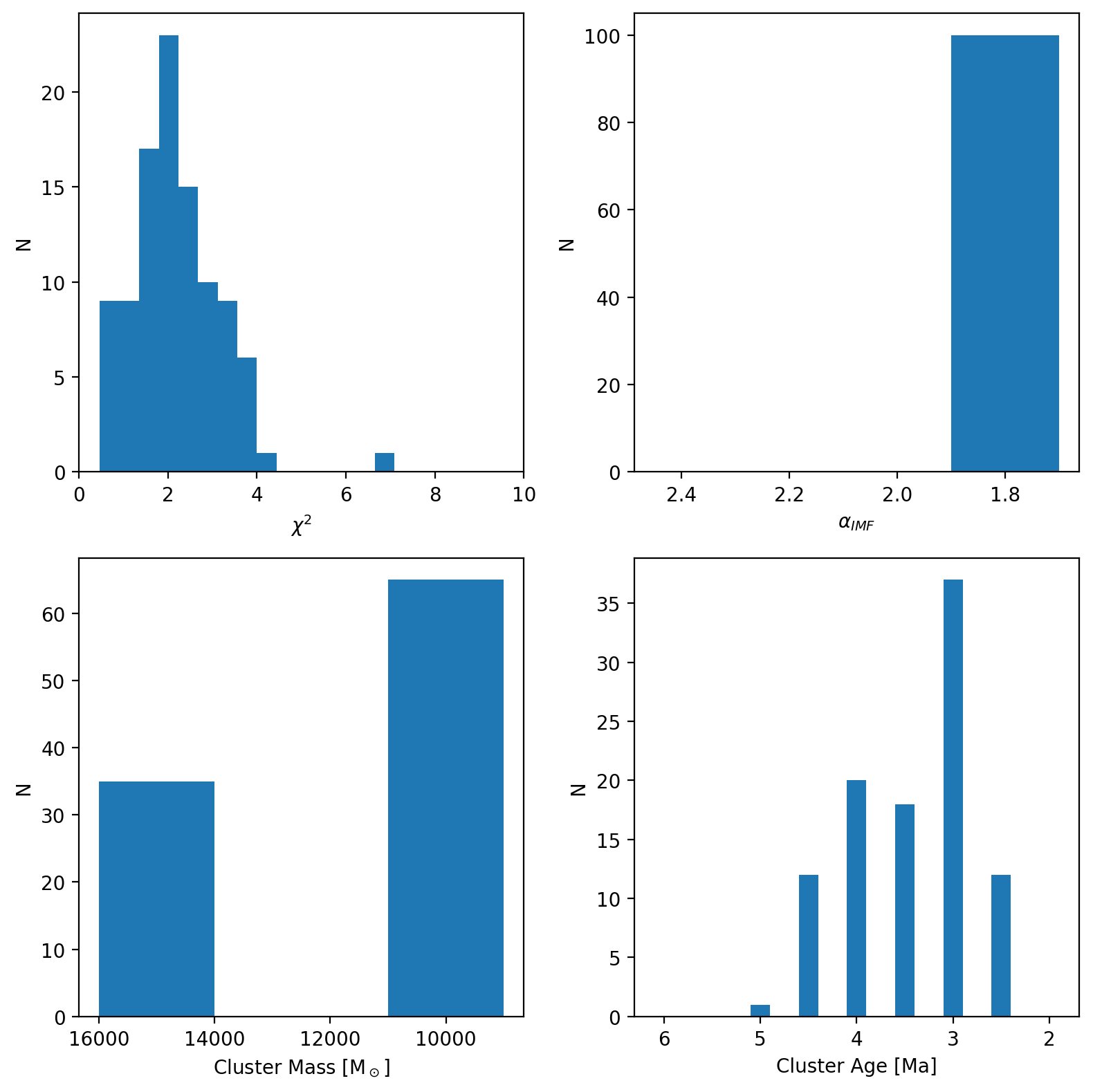}
    \caption[MC simulations of the Arches RLF, using PARSEC isochrones]{Results of MC simulations of the Arches RLF, using PARSEC isochrones.}
    \label{fig:Arches_Parsec}
\end{figure}

We determined the best fit parameters for each of the 100 MC realisations of the RLF observed by \citet{Gallego-Calvente:2021yl}. Figure\,\ref{fig:Arches_MIST} shows the resulting distribution of the parameters of the best fit models with MIST isochrones and Fig.\,\ref{fig:Arches_Parsec} for Parsec isochrones. There is no clear preference for the total cluster mass, but the top-heavy IMF provides the best solution in the vast majority of cases. The cluster age is broadly distributed, but peaks at 3 to 3.5\,Myr, in agreement with literature.

\subsection{Quintuplet cluster}

\begin{figure}[ht!]
    \includegraphics[width=0.5\textwidth]{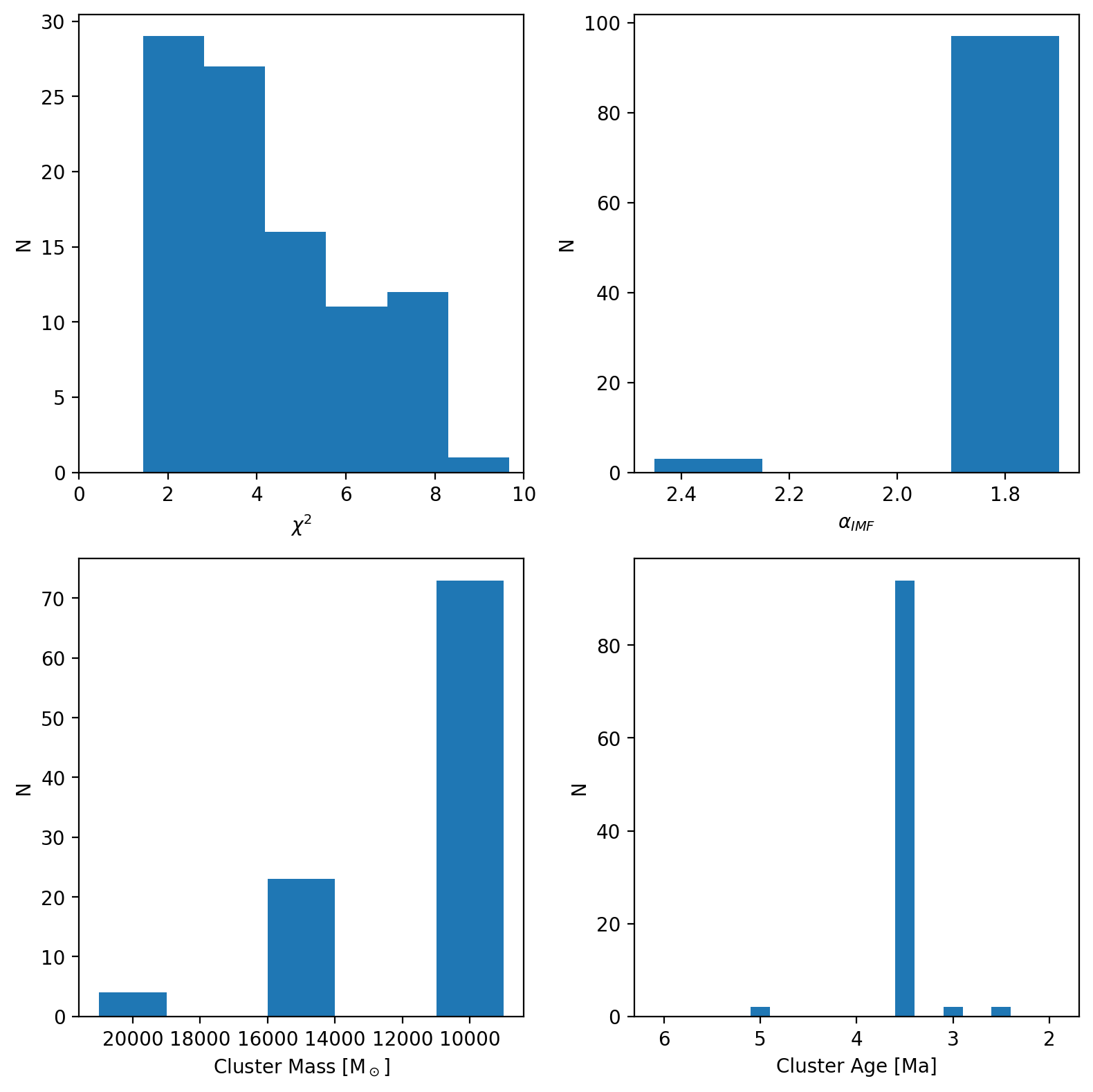}
    \caption[MC simulations of the Quintuplet RLF, using MIST isochrones]{Results of MC simulations of the Quintuplet RLF, using MIST isochrones.}
	\label{fig:Quintuplet_MIST}
\end{figure}

\begin{figure}[ht!]
    \includegraphics[width=0.5\textwidth]{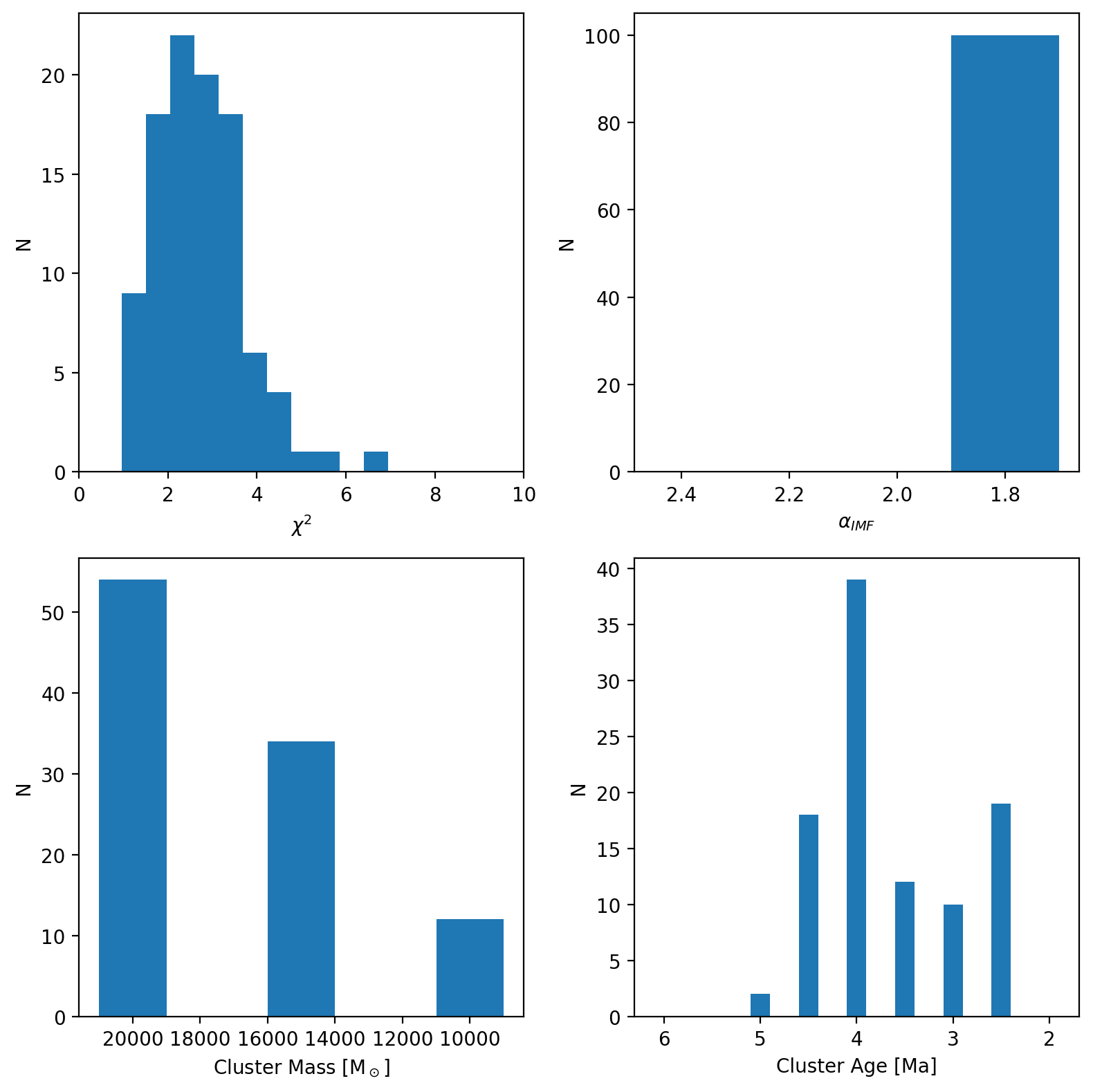}
    \caption[MC simulations of the Quintuplet RLF, using PARSEC isochrones]{Results of MC simulations of the Quintuplet RLF, using PARSEC isochrones.}
    \label{fig:Quintuplet_Parsec} 
\end{figure}

We determined the best fit parameters for each of the 100 MC realisations of the RLF observed by here. Figure\,\ref{fig:Quintuplet_MIST} shows the resulting distribution of the parameters of the best fit models with MIST isochrones and Fig.\,\ref{fig:Quintuplet_Parsec} for Parsec isochrones. As in the case of the Arches cluster, the cluster mass does not appear to be well constrained, but again the top-heavy IMF provides the best solution in the vast majority of cases. The cluster age is broadly distributed and peaks at 3.5 to 4\,Myr, in agreement with literature that assigns an older age to the Quintuplet than to the Arches.  

\section{Discussion}

We have fitted very basic models to MC simulations of the Quintuplet and Arches RLFs. These models:
\begin{itemize}
\item show that theoretical isochrones appear to provide reasonable matches to the observed RLFs of the two clusters;
\item confirm a top-heavy IMF for the two clusters. This agrees with the conclusions of observational studies of the present-day mass functions of these clusters \citep{Husmann:2012vn,Hosek:2019vn};
\item cannot provide any significant constraints on the total cluster mass, but indicate roughly the correct ages of the clusters together with the older age of the Quintuplet cluster.
\end{itemize}

There are important caveats to keep in mind. Above all, not unexpectedly, we find a degree of degeneracy between the IMF slope and the mass of the cluster. Therefore, it is important to constrain the mass and field-of-view (typically the primary beam of the antennas of the array) of the study. This problem is more complicated due to the uncertain detection limit of our radio observations, which suffer from lower S/N and systematics outside of the centre of the FoV. Finally, mass segregation might have concentrated massive stars near the cluster centre, while three-body interactions may have ejected massive cluster members. All these factors increase the systematic uncertainty of this kind of work. Our next step must consist in improved radio observations of the clusters, in order to push down the RLF completeness limit. This can mostly be achieved through a more complete UV coverage of the radio interferometric data by using longer integration times and multi-configuration observations, being sensitive to structures of different angular sizes, to image simultaneously both compact and extended sources. Finally, we can push the radio detections to the limit by using the fact that radio stars must necessarily have bright infrared counterparts. These observations will be very relevant for future SKA observations at mid-frequency (SKA-MID) which would permit the 
study of massive stars and their winds at all stages of evolution up to much lower mass-loss rates, increasing significantly the number of detections.



\begin{acknowledgements}
      Karl G. Jansky Very Large Array (JVLA) of the National Radio Astronomy Observatory (NRAO) is a facility of the National Science Foundation (NSF) operated under cooperative agreement by Associated Universities, Inc.\\
      A.~T. G.-C., R. S., A. A. and B. S. acknowledge financial support from the State Agency for Research of the Spanish MCIU through the ``Center of Excellence Severo Ochoa'' award for the Instituto de Astrof\'isica de Andaluc\'ia (SEV-2017-0709). \\
      A.~T. G.-C., R. S., and B. S. acknowledge financial support from national project PGC2018-095049-B-C21 (MCIU/AEI/FEDER, UE).\\
      A. A. acknowledges support from national project PGC2018-098915-B-C21 (MCIU/AEI/FEDER, UE). \\
      F. N. acknowledges financial support through Spanish grants ESP2017-86582-C4-1-R and PID2019-105552RB-C41 (MINECO/MCIU/AEI/FEDER) and from the Spanish State Research Agency (AEI) through the Unidad de Excelencia ``Mar\'ia de Maeztu''-Centro de Astrobiolog\'ia (CSIC-INTA) project No. MDM-2017-0737. \\
      F. N.-L. gratefully acknowledges funding by the Deutsche Forschungsgemeinschaft (DFG, German Research Foundation) -- Project-ID 138713538 -- SFB 881 (``The Milky Way System'', subproject B8).\\
      F.N.-L. acknowledges the sponsorship provided by the Federal Ministry for Education and Research of Germany through the Alexander von Humboldt Foundation.\\
      The research leading to these results has received funding from the European Research Council under the European Union's Seventh Framework Programme (FP7/2007-2013) / ERC grant agreement n$^{\circ}$ [614922].
\end{acknowledgements}

\bibliographystyle{aa} 
\bibliography{Quintuplet.bib} 

\end{document}